\documentclass[preprint]{aastex}

\shortauthors{Budav\'ari {\it et al.}}
\shorttitle{Spectral Templates from Photometry}
\newcommand{\etal}{\it et al.}

\begin{document}
\title{Creating Spectral Templates from Multicolor Redshift
Surveys\altaffilmark{1}} 

\author{Tam\'as  Budav\'ari\altaffilmark{2}, Alexander S.  Szalay}
\affil{Department of Physics and   Astronomy, The Johns  Hopkins  University,
Baltimore, MD 21218}
\email{budavari@pha.jhu.edu}

\author{Andrew J. Connolly\altaffilmark{3}}
\affil{Department of Physics and Astronomy, University of
Pittsburgh, Pittsburgh, PA 15260}

\author{Istv\'an Csabai\altaffilmark{3}}
\affil{Department of Physics, E\"{o}tv\"{o}s University, Budapest,
Pf. 32, Hungary, H-1518}

\and
\author{Mark Dickinson}
\affil{Space Telescope Science Institute, Baltimore, MD 21218}

\altaffiltext{1}{Based on observations with the NASA/ESA
Hubble Space Telescope, obtained at the Space Telescope Science Institute,
which is operated by the Association of Universities for Research in
Astronomy, Inc., under NASA contract NAS5-26555.}
\altaffiltext{2}{Department of Physics, E\"{o}tv\"{o}s University, Budapest,
Pf. 32, Hungary, H-1518} 
\altaffiltext{3}{Department of Physics and   Astronomy, The Johns  Hopkins
University, Baltimore, MD 21218}

\begin{abstract}
Understanding how  the physical properties of galaxies  (e.g.  their spectral
type or  age) evolve as a function  of redshift relies on  having an accurate
representation of  galaxy spectral energy  distributions.  While it  has been
known for some  time that galaxy spectra can be  reconstructed from a handful
of orthogonal  basis templates, the  underlying basis is  poorly constrained.
The limiting factor has been the  lack of large samples of galaxies (covering
a wide  range in spectral type) with  high signal-to-noise spectrophotometric
observations. To alleviate this problem we introduce here a new technique for
reconstructing galaxy spectral energy  distributions directly from samples of
galaxies  with  broadband   photometric  data  and  spectroscopic  redshifts.
Exploiting the  statistical approach  of the Karhunen-Lo\`eve  expansion, our
iterative training procedure increasingly improves the eigenbasis, so that it
provides better agreement with the photometry.  We demonstrate the utility of
this approach by applying these improved spectral energy distributions to the
estimation of photometric  redshifts for the HDF sample  of galaxies. We find
that  in a  small  number of  iterations  the dispersion  in the  photometric
redshifts estimator  (a comparison between predicted  and measured redshifts)
can decrease by up to a factor of 2.
\end{abstract}


\section{Introduction} \label{intro}

With the introduction of large-format  CCDs, the study of galaxy evolution is
undergoing a  renaissance. Statistically significant volumes  of the Universe
can now be surveyed at high and low redshift with relatively small amounts of
observing time.  In fact the rate  at which we can undertake deep, multicolor
photometric surveys is now over 2 orders of magnitude larger than the rate at
which we  can followup these observations spectroscopically.   As an example,
the  Sloan Digital  Sky  Survey will  spend  85\% of  its  five year  mission
measuring the  redshifts of  one million objects  with a  unique, multi-fiber
spectrograph, while in the remaining 15\%  of the time it will gather 5 color
photometry for 100 million objects.

This  rapid  growth in  multicolor  surveys has  led  to  the development  of
techniques for  exploiting the  information contained within  the photometric
data themselves (without  relying on followup spectroscopy). One  of the most
successful of  these techniques has  been the estimation of  galaxy redshifts
directly from  multicolor data (photometric  redshifts) \citep{baum62}.  (For
an  overview of  recent work  in this  field, see  \citet{weymann99}.)  These
techniques can be  broken down into two basic classes.   One class of methods
use a training set of  galaxies with known photometry {\it and} spectroscopic
redshifts to derive an empirical  relation between the colors of galaxies and
their redshifts \citep{connolly95a, wang98}.  After calibration, the redshift
estimator   can  predict   redshifts  for   objects  with   only  photometric
observations. The advantage of this  method lies in its empirical nature; all
the effects of dust and galaxy evolution that are present in the training set
are implicit  within the  derived correlation.  The  drawback is that  it can
only robustly predict redshifts for objects  that are similar to those in the
original  training   set.   In  other   words  the  estimator  is   good  for
interpolation, but not  for extrapolation to galaxies that  have redshifts or
spectral types far from the range covered by the training set.

The  second   class  of  techniques,  template   based  photometric  redshift
estimators (or spectral energy  distribution fitting), use model or empirical
galaxy  spectral  energy  distributions  \citep{koo85,  connolly95b,  gwyn96,
soto99, benitez99}.   By minimizing the  difference between the  observed and
synthetic  colors one  can find  the  most likely  type and  redshift of  the
galaxy. This method is simple to  implement, does not require a training set,
and  has  no  extrapolation  problems.   Its  main  limitation  is  that  the
underlying spectral energy distributions (SEDs) of galaxies within the sample
must  be well known.   Comparisons of  the colors  of galaxies  with spectral
synthesis models \citep{bc93} have shown that the modeling of the ultraviolet
spectral  interval  for  galaxies  is  uncertain.  Whether  this  is  due  to
uncertainties in the modeling  of the stars or due to the  effects of dust is
as  yet  unclear.   Consequently,  photometric-redshift  estimates  are  most
accurate when  we apply empirical spectral energy  distributions derived from
observations  of local  galaxies, e.g.   \citet{sawicki97}.   These empirical
relations are,  however, constructed  from a small  number of  local galaxies
that  have been  observed  in detail  and  there is  no  guarantee that  they
represent the full distribution of galaxy types (particularly when we include
the effects of evolution with redshift).

In this  paper we derive a method  that combines the advantages  of the above
techniques without their drawbacks. We use a training set, not for deriving a
direct  relationship between  colors and  redshifts but  instead to  build an
optimal set of spectral templates. We  optimize for the shape of the spectral
templates  to  give  the  best  match between  the  predicted  galaxy  colors
(calculated  using  the  {\it  spectroscopic}  redshifts)  and  the  observed
colors. As a result  we derive a set of spectral templates  that are a better
match  to the  SEDs of  the galaxies  in the  training set  than  the initial
model/empirical templates. Our tests  show that these improved templates give
a  tighter photometric  redshift relation  than  do the  original SEDs.

In  section   \ref{inversion}  we  revisit   different  photometric  redshift
estimator  techniques  and  we  address   the  question  of  how  to  recover
information  beyond just redshift  from photometric  observations and  how to
interpret  these results. We  introduce an  interpolation scheme  for optimal
spectral  type determination.   In  section \ref{algorithm}  we describe  our
novel technique for creating eigenspectra in an iterative way, and in section
\ref{app} we show a test application to the HDF-N/NICMOS catalog, where there
are around 150 spectroscopic  redshifts available.  We present the comparison
of photometric redshift estimates based on the most commonly used \citet{cww}
spectra (hereafter CWW) and our empirically developed templates. 
 
\section{Photometric redshifts} \label{inversion}

Estimating the  redshift of  a galaxy from  broadband photometry relies  on a
comparison between the colors of the galaxy and those predicted from either a
set  of  spectral  energy  distributions  (SEDs) or  an  empirically  derived
relation between color  and redshift. The accuracy of  the resulting relation
relies critically  on having a  well calibrated relation (e.g.\  the spectral
templates  must cover  the full  range  of galaxy  types). In  this paper  we
describe a new technique for developing  a set of spectral templates that are
optimized  to the  data  in hand.  Initially,  we outline  a  variant on  the
standard  template  fitting  technique  that  uses  an  orthogonal  basis  to
compactly describe the broad range of galaxy spectral types.

\subsection{Data} \label{data}

Unique photometric measurements  in four bands (U300, B450,  V606, I814) have
been available from the Hubble  Space Telescope's Wide Field Planetary Camera
2  (WFPC2) since  1996, when  a  slice of  the very  distant Universe  became
visible in  the Hubble Deep  Field (HDF-N)\citep{williams96}.  This  data set
has been the  testbed of photometric redshift estimators  from the beginning.
Recently the HDF-N field gained  two more high signal-to-noise infrared bands
(J110,  H160) when  the Near  Infrared Camera  and Multi  Object Spectrometer
(NICMOS) targeted  the same area \citep{dickinson00}.   The joint photometric
catalog  of  the  observations   was  extracted  by  the  SExtractor  package
\citep{sextractor} using  a weighted  sum of  the J and  H NICMOS  images for
object detection.  In the larger  view of the NICMOS instrument, 1681 objects
were detected,  and out of these,  1291 objects were  extracted with complete
photometric information  in all six bands.   In addition to the  HST data, we
have used  $K_s$--band photometry  derived from images  taken at the  KPNO 4m
telescope  \citep{dickinson98}.  We  have used  a method  similar to  that of
\citet{soto99}  for reliably extracting  fluxes and  errors for  every object
detected  in  the   NICMOS+WFPC2  images  \citep{papovich00}.   This  ensures
photometric  completeness at  $K_s$ and  eliminates  uncertainties concerning
aperture corrections between the  HST and ground--based data sets.  Detection
in  both optical  and  infrared enables  photometric  redshift estimators  to
identify very  high redshift objects out  to about $z=13-14$,  because of the
continuum blanketing from Lyman-$\alpha$ forest \citep{madau95}.

The  number of  spectroscopic redshifts  in HDF-N/NICMOS  field is  also very
promising for various astrophysical  projects.  Currently there are about 150
spectroscopic  redshifts  available   for  the  central  HDF  \citep{cohen96,
steidel96, lowenthal97, dickinson98,  hogg98, weymann98, spinrad98, stern99}.
These spectroscopic measurements  can be used as a  training set in empirical
photometric redshift estimations or as a test of the accuracy and reliability
of template fitting algorithms.

\subsection{Using eigenspectra} \label{eigen}

Traditional SED fitting methods provide a simple mechanism for estimating the
redshift and  spectral type (or  metallicity, age etc)  of a galaxy.   One can
compare the measured and the template  based colors as a function of redshift
and select  the redshift and the template  where the match is  the best.  The
limitation of this  method is that it  does not allow a true  estimate of the
uncertainties  on these  measures of  age etc.,  nor does  it  facilitate the
propagation   of  these  errors   to  subsequent   analyses  of   the  galaxy
distributions.  The  reason for  this is that  the multidimensional  space of
parameters that describe a galaxy are only sampled over a few discrete points
(i.e. the number of template spectra).

Instead of using a  discrete set of SED templates one can  create from them a
continuous  manifold.   We  can  imagine   SED  templates  as  vectors  in  a
multi-dimensional  vector space;  the dimension  of this  space ($D$)  is the
number of  the wavelength bins of  the SEDs. With an  orthogonal basis within
this vector  space one  can expand  any SED, $S(\lambda)$,  in terms  of this
basis,
\begin{equation}
S(\lambda) = \sum_{j=1}^{D} c_j \, \Psi^j(\lambda)
\label{fullEigen}
\end{equation}
where $\Psi^j(\lambda)$ is the $j$th  basis spectrum and $c_j$ are the linear
expansion coefficients.

One  ends up  with  $D$ coefficients  to define  a  SED instead  of a  single
discrete spectral type parameter that selects the template in the traditional
way. In practice,  the SEDs do not have arbitrary shapes:  they occupy only a
small a subspace of the possible $D$-dimensional vector space.  With the help
of  the the Karhunen-Lo\`eve  (KL hereafter)  transformation \citep{karhunen,
loeve, connolly95b, csabai00}, a variant of the principal component analysis,
we  can systematically  define a  small number  of basis  vectors ($\Psi^j$),
which can reproduce  all real spectra to a high  (and quantifiable) degree of
accuracy.   Using  only   $\eta$  eigentemplates  the  approximated  spectrum
$\hat{S}(\lambda)$ can be expressed in the form of
\begin{equation}
\hat{S}(\lambda) = \sum_{j=1}^{\eta} c_j \, \Psi^j(\lambda)
\label{approxEigen}
\end{equation}
where  the  value of  $\eta$  is  typically $2$  to  $4$.   We  call this  an
eigenbasis, and its vectors are the  eigenspectra.  In this way the number of
free parameters for which we have  to optimize remains small, but we have the
freedom of having a  continuous distribution of template spectra. Photometric
redshift estimation  follows the same  approach of comparing the  derived and
actual colors but  the coefficients of the eigenspectra are  saved as well as
the redshift.  

Truncating  the original  basis introduces  an uncertainty  as a  function of
wavelength  into  the linear  combination  (see equation  \ref{approxEigen}),
which can expressed in terms of the remaining eigenspectra as
\begin{equation}
\sigma^2(\lambda) = 
  \sum\limits_{j=\eta+1}^{D} \Lambda_j \Big( \Psi^j(\lambda) \Big)^2
\label{errEigen}
\end{equation}
where $\Lambda_j$ is the eigenvalue of the $j$th eigenspectra.

Traditionally    the    eigentemplates    are    derived    from    empirical
\citep{cww,kinney96} or  synthetic \citep{bc93} spectra.   Unfortunately both
methods  have  limitations  when  it  comes to  applications  to  real  data.
Empirical spectra are  typically derived from observations of  a small number
of local galaxies. The derived basis  may not, therefore, be tenable for high
redshift galaxies.  A second problem is  that the there is only a small range
in  wavelength where  good signal-to-noise  data  can be  obtained; e.g.\  UV
observations  or  IR  spectrophotometry   of  local  galaxies  are  extremely
limited. In contrast, synthetic spectra  are available over a wide wavelength
range and can  be derived for galaxies  with a wide variety of  ages and past
star formation  histories, yet many previous  studies, e.g. \citet{sawicki97,
yee98, benitez99},  have found  that they do  not yield  photometric redshift
estimates as  reliable as those  derived from empirical  templates.  Reliable
templates  could  greatly  enhance   the  range  where  photometric  redshift
estimation techniques can be applied with acceptable errors.

The  eigenbasis approach  is  more  sophisticated in  several  ways than  the
traditional discrete  template fitting  version.  As described  previously, a
small number  of continuous parameters  are used to determine  the rest-frame
spectrum, allowing proper error propagation.  These parameters are the linear
coefficients of  the eigenspectra  and their combination  can be  related not
only to a single spectral type  parameter (as in the traditional version) but
also, in  principle, to other physical  parameters such as  reddening, age of
the galaxy etc.

While there remains no standard mechanism for defining the spectral type of a
galaxy  from  the  eigentemplates,  it  has  been shown  that  two  or  three
eigentemplates can  yield reasonably  accurate approximations to  most normal
galaxy spectra \citep{connolly95b}.  Both the  luminosity of a galaxy and its
spectral type  are encoded  within these coefficients.  In this study  we use
three  eigenspectra.  Extracting  the  apparent luminosity  from  the  linear
coefficients  is  straight  forward  (e.g.\  the length  of  the  coefficient
vector).  The  normalized Cartesian  coordinates can then  be converted  to a
polar system  where the spectral type  is defined by two  mixing angles.  The
solid line in figure \ref{mix} shows  these mixing angles as defined from the
CWW spectral  energy distributions.  The  galaxy spectral types are  found to
lie on a one parameter sequence within this two dimensional space.

The points  in figure \ref{mix}  show the mixing  angles as defined  from the
HDF-N  WFPC2 and  NICMOS  data. The  ellipses  are indicative  of the  errors
associated with the  derived mixing angles.  Many of  the galaxies within the
HDF-N data  set are seen  to follow the  locus defined by the  CWW templates.
There are however a number of outlier data points (with correspondingly large
error  ellipses). These  outliers are  typically high  redshift  objects with
reasonably accurate  predicted redshifts  but with poorly  estimated spectra.
What happened?   The rest-frame  spectrum of a  high redshift object  is only
sampled  at short  wavelengths, thus  the cost  function in  the optimization
measuring  the deviation  of  the actual  and  template based  colors is  not
sensitive to the  longer wavelength end. In other words  the cost function is
``flat'' in the expansion coefficient  subspace, so tiny deviations drive the
resulting spectral type,  while the cost vs.  redshift  has a normal minimum.
This observation enables us to  introduce a method to avoid negative spectra,
while also reducing the number of fitted parameters.

\subsection{Optimal subspace filtering} \label{optimal}

We propose a physical interpolation  scheme to resolve the type degeneracy of
high redshift  objects and  to insure plausible  spectra for any  galaxy.  We
derive  a one-dimensional type  parameter.  As  a starting  point we  use the
empirical  CWW spectra, extrapolated  in the  ultraviolet and  infrared using
empirical  templates  from  \citet{kinney96}  and models  from  \citet{bc93},
respectively, and with the addition  of a model SED \citep{bc93} representing
a very  blue star forming  galaxy.  We then  use these to derive  our initial
eigenbasis, truncated  to a  3-dimensional subspace.  The  KL expansion  of a
spectrum  on  this eigenbasis  gives  coefficients  that  represent a  galaxy
spectrum.  We found  that the normalized coefficients of  the CWW spectra are
in a region close to a small  circle on the unit sphere.  This region defines
a one-dimensional sequence, that can be mapped to the $[0,1]$ interval, where
the value $0$  roughly corresponds to ``Ellipticals'' and  $1$ represents the
bluest galaxies.   Deriving the  above {\it type}  parameter consists  of the
following  steps.   In the  3D  coefficient space,  fitting  a  plane to  the
normalized coefficient vectors computed for the reference (CWW) spectra gives
the equation of  a cone, which can be parameterized  with one variable.  This
periodic variable is restricted to the physically sensible interval where the
non-negative spectra  are.  Choosing ``0''  for the best fit  ellipticals and
``1'' for the bluest galaxies  yields an easy interpretation of the estimated
type.

\begin{figure}
\epsscale{0.85}
\plotone{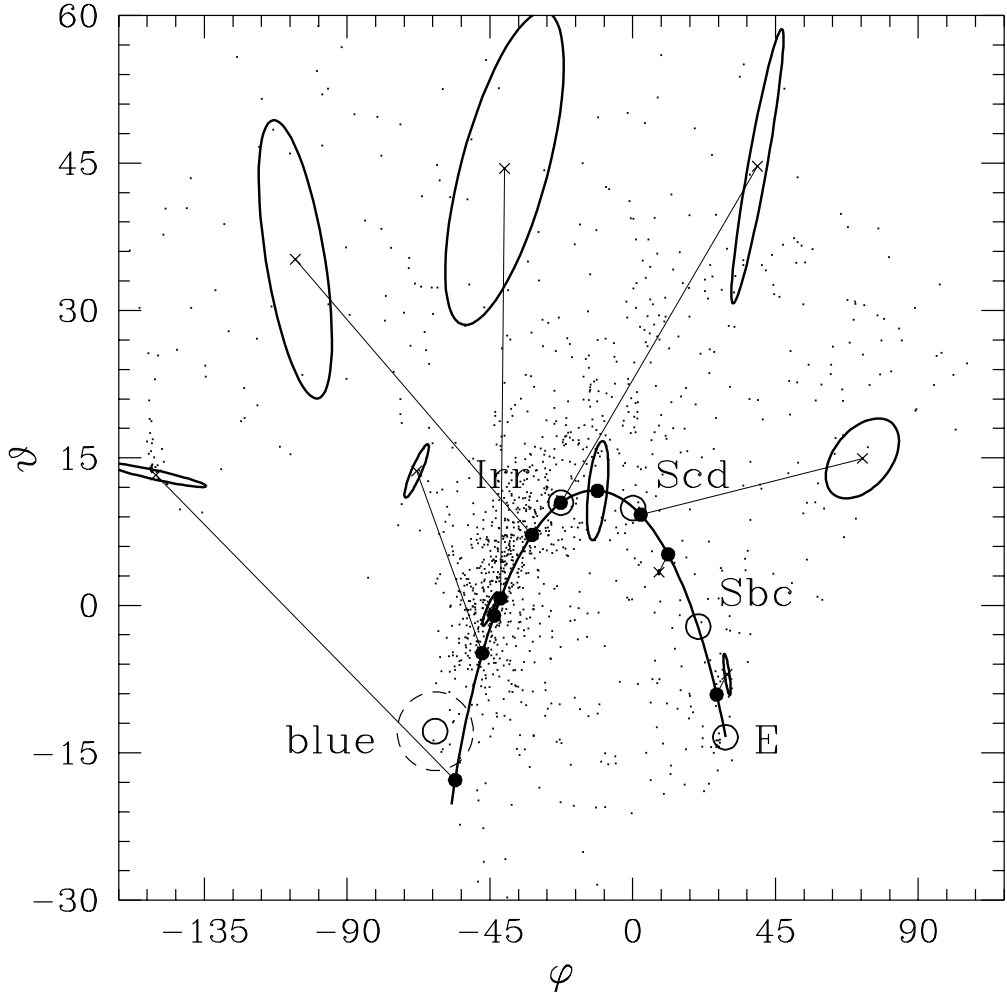}
\caption{Using three eigentemplates, the spectral type
of a galaxy can be parameterized with two mixing angles in polar coordinates,
$\varphi$ and $\vartheta$.  Having projected a spectrum to the eigenbasis,
one can determine the corresponding $(\varphi,\vartheta)$ pair. The
coordinates calculated for the CWW spectra are plotted in the figure with
open circles.  Points represent estimated spectra computed applying the
standard eigenspectra method (see section \ref{eigen}) and their errors are
also shown for some random sample galaxies with open ellipses. The solid
curve illustrates the interpolation scheme as introduced in section
\ref{optimal}, the CWW locus.  In the constrained fit, varying the
one-dimensional {\it type} parameter and mapping it analytically to the
mixing angle space guarantee physically sensible spectra. Thin lines
connecting the center of the error ellipses to the 1D subspace illustrate how
the estimated continuum spectra change when moving from the pure
eigenspectrum fitting to the constrained type fitting.  A dashed circle is
also drawn around the blue reference template to emphasize that it was taken
from a synthetic catalog (Bruzual \& Charlot 1993) and was not used in the
type sequence derivation. \label{mix}}
\end{figure}

Exploiting the  above definition  of a scalar  type parameter, we  develop an
improved redshift and type estimator \citep{connolly99}.  In this application
there are three continuous fitting parameters: redshift, type and luminosity.
In  figure \ref{mix}  open circles  represent the  mixing angles  of  the CWW
spectra and the  solid line shows the projected trajectory  from one CWW type
to the next.   The points in figure \ref{mix} show the  mixing angles for the
HDF  sample and  the  ellipses  illustrate the  corresponding  errors on  the
derived angles.  The type and redshift  errors are computed from the shape of
the cost  function around  the minimum, and  the formal covariance  matrix is
obtained  from the  coefficients of  a fitted  paraboloid. Looking  at figure
\ref{mix},  one can  see that  the error  ellipses on  the mixing  angles are
larger as we move farther from the  CWW locus and that the position angles of
these error  ellipses point in a  direction that is almost  orthogonal to the
CWW curve.

The effect of the optimal subspace filtering is to force the mixing angles to
reside on the  CWW trajectory. In figure \ref{mix}  thin lines connect sample
objects with  error ellipses to  the corresponding constrained  mixing angles
lying  on   the  CWW  locus.   This   physically  motivated  model---smoothly
interpolating  between the  CWW SEDs---provides  sensible spectral  types for
higher redshift  galaxies by suppressing the  effect of noise  on the derived
expansion  coefficients.  Simply  using only  the  discrete set  of four  CWW
templates would clearly result in unrealistic estimated types for most of the
galaxies, eventuating false photometric redshifts.

\section{Template reconstruction algorithm} \label{algorithm}

We have seen previously  that a small number of basis spectra  can be used to
approximate any spectrum reasonably well. The limitation at this point is the
quality of the libraries that the eigentemplates are built from.  In order to
improve the accuracy of the photometric redshifts and spectral typing we need
to build  better eigenspectra.  The  first step was \citet{csabai00}  using a
direct technique to optimize for the shape of the eigenspectra simultaneously
with  the coefficients.   Here  we follow  a  different statistical  approach
allowing  a higher  wavelength resolution  even for  the  currently available
small data sets \citep{budavari99}.

\subsection{The idea: spectra from broadband photometry} \label{idea}

\begin{figure}
\epsscale{0.85}
\plotone{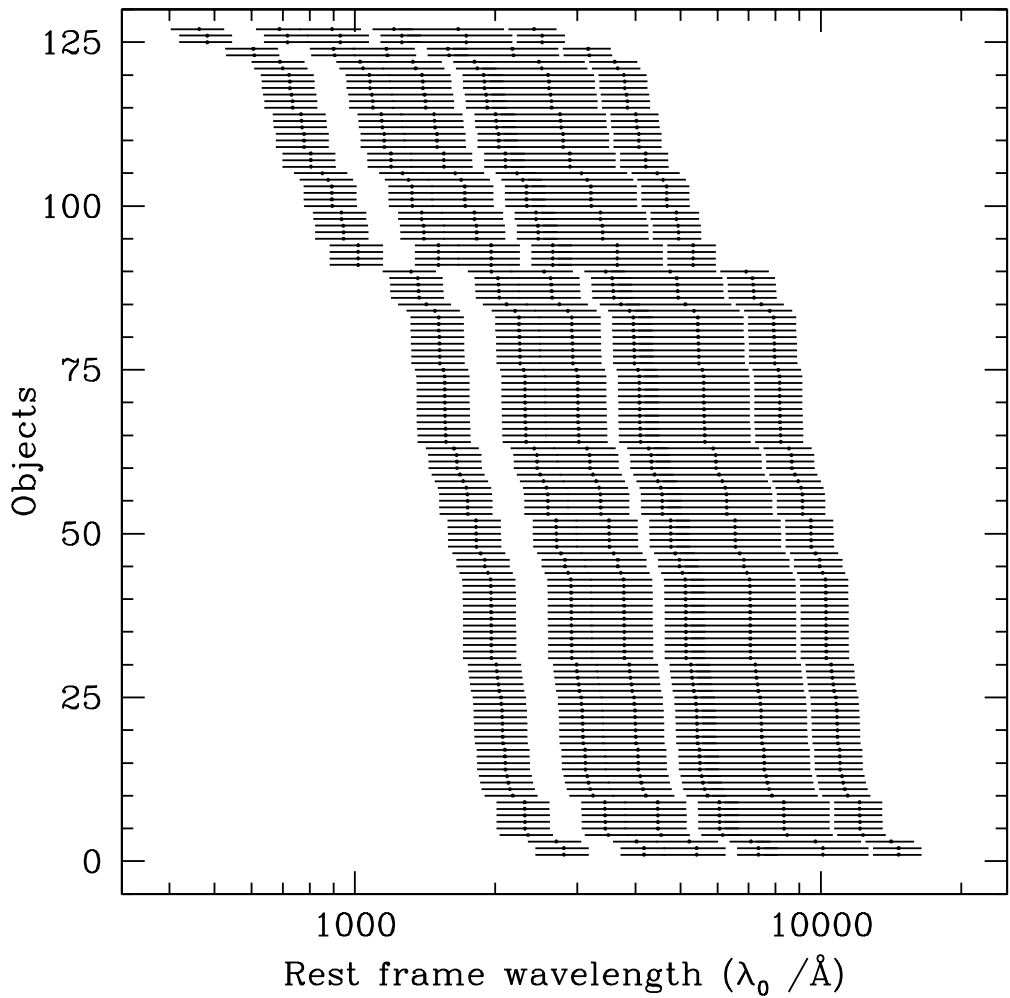}
\caption{For a sample of objects spanning a range of redshifts, the rest frame 
wavelengths covered by a set of fixed photometric bandpasses will overlap 
when projected back to the rest frame.  This is illustrated here using a set 
of HDF--N galaxies with spectroscopic redshifts.  The solid bars show the 
rest frame wavelength ranges covered by the WFPC2 + NICMOS filters.
\label{seds}}
\end{figure}

How can we obtain spectra  from broadband photometry?  We observe galaxies at
different redshifts, which  means the filters sample the  rest-frame light at
different wavelengths.   If we have  large multicolor redshift  surveys, then
the  wavelength   regions  sampled  by   the  filters  (in   the  rest-frame)
overlap. Figure  \ref{seds} illustrates how the  HDF/NICMOS photometric bands
(projected  back  to the  rest-frame)  overlap  for  a set  of  spectroscopic
redshifts taken from the catalog.  Looking at a particular wavelength, we can
actually see which bands set up constraints on the value of the eigenspectra.
This oversampling  of the eigenspectra  allows us to develop  applications to
extract  higher  resolution information.   Developing  such a  reconstruction
process  is not  trivial, though.   Several  problems emerge  because of  the
non-uniform  redshift distribution and  the photometric  uncertainties within
the data.  Obviously none of  the standard de-convolution applications can be
used  to solve  the  problem.   We have,  therefore,  developed an  iterative
procedure for  correcting eigenspectra,  which repairs the  eigenspectra only
where  sufficient   information  exists  and  where  the   magnitude  of  the
modification is determined statistically.

Our  iterative procedure  starts from  an initial  set of  eigentemplates and
creates  increasingly better eigenbases  by improving  the match  between the
measured colors and those derived from the templates.  The algorithm is quite
simple.   Given   a  catalog  of  galaxies  with   broadband  photometry  and
spectroscopic redshifts, we first expand  each galaxy over our initial set of
templates, transformed to the  galaxy's redshift, by  solving for  the linear
expansion coefficients.

If the initial templates do not accurately reproduce the ensemble of galaxies
the spectrum of every galaxy will have systematic errors due to the choice of
templates.   This corresponds  to a  subspace in  the space  of  all possible
spectra.  A better representation of the photometry may require moving out of
this subspace.  We  can achieve this by an iterative  repair of the estimated
spectrum of  each galaxy, by  considering what is  the minimal change  in its
shape  that would  improve the  photometry.   We compute  this ``repair''  by
minimizing a  cost-function over the  shape of the spectrum,  penalizing both
large deviations from the photometry  and from the template expansion.  Since
the  template  expansion was  already  an  optimal  representation given  the
subspace defined  by the templates, this  repair enables the  spectra to move
out of the original subspace.

\begin{figure}
\epsscale{0.85}
\plotone{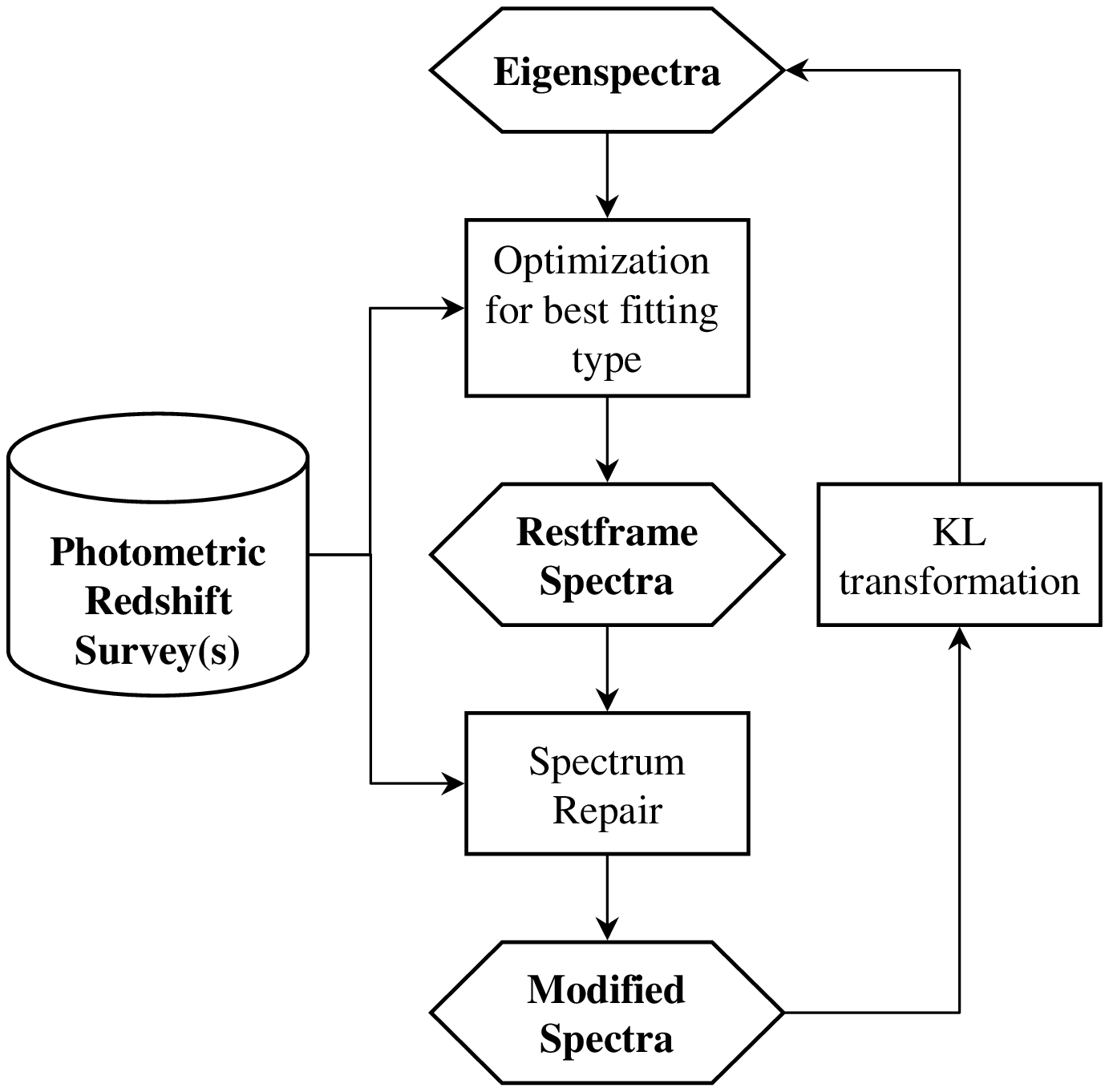}
\caption{The reconstruction process is an iterative
training algorithm. Starting from a set of eigenspectra, a best fitting
spectral type can be derived for each object in a catalog from the
spectroscopic redshift and photometric information. The type determines the
rest-frame spectrum via the linear combination of the eigentemplates. Having
``repaired'' these spectra, a new generation of eigenspectra derived by
applying the KL expansion. \label{flow}}
\end{figure}

Building a new eigenbasis over the set of repaired spectra will allow changes
on the templates which improve the fit population-wide, but reject individual
fluctuations.  Performing this process multiple times in an iterative fashion
\citep{gappykl},    the   overall    fit   to    the    photometry   improves
substantially. The flowchart of the algorithm is shown in figure \ref{flow}.

\subsection{Repairing spectra} \label{repair}

The  key  step to  the  template  reconstruction  algorithm is  the  spectrum
repairing method. The spectrum we are  about to repair is an approximation of
the real spectrum based on  the photometric data.  This best fitting spectrum
can  be found  within the  subspace spanned  by the  current  eigenspectra by
varying the  coefficients directly. To obtain a  physically sensible spectrum
the optimal  subspace filtering method (see section  \ref{optimal}) should be
used.  Having  a set of eigentemplates  and the calculated  best fitting type
the corresponding  rest-frame spectrum  of a galaxy  can be  easily computed.
The best fitting type $t$ is calculated by minimizing the cost function,
\begin{equation}
C^2(t) = \sum\limits_n {1\over \Delta_n^2}
             \Big( \hat{f}_n - f_n(t,z_{\rm spec}) \Big)^2
\end{equation}
where $\hat{f}_n$,  $\Delta_n$ are the measured  flux and error  in the $n$th
passband.   In fact,  the type  parameter $t$  determines the  values  of the
coefficients of  the eigenspectra and  the linear combination gives  the best
fitting  spectrum.  Using  $\eta$ number  of eigentemplates  the approximated
spectrum can be expressed in the form of
\begin{equation}
\hat{S}(\lambda) = \sum_{j=1}^{\eta} c_j(t) \, \Psi^j(\lambda) \label{hatS}
\end{equation}
where  $\Psi^j(\lambda)$ is  the  $j$th eigenspectrum  and  $c_j(t)$ are  the
linear coefficients defined by the type $t$ and the brightness.

The  basic idea  is to  modify  this spectrum  to match  the observed  colors
better.  This  is another  minimization problem, where  we try to  adjust the
spectrum based  on a cost function built  up by two different  terms.  On one
hand, we have the deviation of the actual photometry from the fluxes computed
from  the spectrum  itself,  $S(\lambda)$.  On  the  other hand  we have  the
deviation  of   the  spectrum   $S(\lambda)$  from  the   template  expansion
$\hat{S}(\lambda)$.  The repair process has  to be able to deal properly with
different  types   of  errors  such   as  the  photometric  errors   and  the
uncertainties within the eigenspectra.

Let  $R^{n}(\lambda)$ represent the  response function  of the  $n$th filter,
normalized and corrected for  the detector and telescope throughput function.
The  derived flux in  the $n$th  passband can  be written  as an  integral or
approximated with the sum over the discrete representation of the functions.
\begin{equation}
f_n = \int R^{n}(\lambda) \, S(\lambda) \, d\lambda = \sum_k r^{n}_k
       \, s_k
\end{equation}
where  $s_k$ and  $r^n_k$ are  the  discrete representation  of the  spectrum
$S(\lambda)$  and the  filter  $R^n(\lambda)$  and $k$  refers  to a  certain
wavelength, $\lambda_k$.  The discrete version of $\hat{S}(\lambda)$ is given
by the linear combination of  $c_j(t)$ and $\psi^j_k$, the new representation
of the $j$th eigenspectrum $\Psi^j(\lambda)$.

The discrete  problem translates into a multidimensional  minimization of the
cost function as a function of the unknown spectral shape $s_k$.
\begin{equation}
\chi^2 = \sum_k { 1 \over \sigma_k^2 } \Big( s_k - \hat{s}_k \Big)^2
	+ \sum_n { 1 \over \Delta_n^2 } \Big( f_n - \hat{f}_n \Big)^2
\end{equation}
where $\sigma_k$ describes the ability of the eigenspectra to be changed at a
certain  wavelength.  The primary  source of  error in  the expansion  is the
truncation error due to  the restricted subspace; see equation \ref{errEigen}
and its discrete form,
\begin{equation}
\sigma_k^2 =\sum_{j>\eta} \Lambda_j \, \Big( \psi^j_k \Big)^2 .
\end{equation}

\begin{figure}
\epsscale{0.85}
\plotone{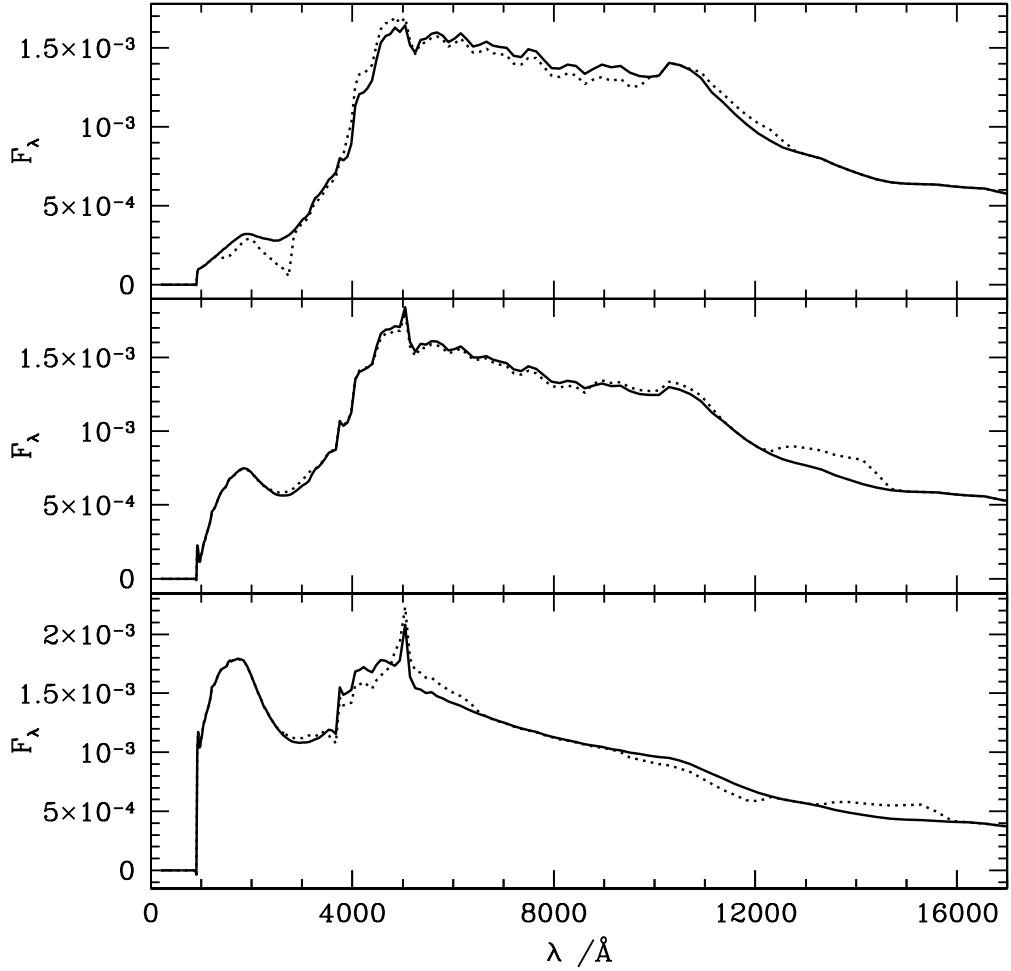}
\caption{The key step of the training algorithm is the
modification of the spectra to improve the correlation of the measured and
template based colors. The modified spectra (dotted line) can be compared to
the originals (solid line) for sample objects. The spectra are normalized to
unit length, the dot product of the spectrum with itself is unity. The types
(see section \ref{optimal}) of these galaxies are 0.06 (E/S0), 0.15 (Sbc) and
0.33 (Scd) respectively from the top to the bottom. \label{modspec}}
\end{figure}

The cost  function of the problem can  be transformed into a  simpler form by
introducing the difference spectrum ($\xi$) and some constant terms ($g_n$).
\begin{eqnarray}
\xi_k & = & s_k - \hat{s}_k \\
g_n & = & \hat{f}_n - \sum_k r^{n}_k \, \hat{s}_k
\end{eqnarray}
\begin{equation}
\chi^2 = \sum_k { \xi_k^2 \over \sigma_k^2 }
	+ \sum_n { 1\over \Delta_n^2}
	\Big(g_n-\sum_k r^{n}_k \, \xi_k \Big)^2
\end{equation}
The cost function is quadratic in its variables, $\{\xi_k\}$. The minimum can
be  analytically  derived,  since  the  partial  derivatives  vanish  at  the
minimum. This  yields a system of  linear equations, which can  be solved for
$\xi$.
\begin{equation}
{ \partial \chi^2 \over \partial \xi_l } = 0 \ ; \qquad
\sum_k M_{lk} \, \xi_k = \nu_l
\end{equation}
where
\begin{equation}
M_{lk} = \sum_n { r^{n}_l \, r^{n}_k \over \Delta_n^2} + {\delta_{lk}
		\over \sigma_k^2} 
\ \  {\rm and} \ \ 
\nu_l = \sum_n { r^{n}_l \, g_n \over \Delta_n^2 }
\end{equation}
and $\delta_{lk}$ is  the usual Kronecker symbol.  The  repaired spectrum can
be computed  from the difference vector  by simply adding it  to the template
based spectrum,  $s_k = \hat{s}_k +  \xi_k$.  A few  typical example repaired
spectra  can be  seen  in  figure \ref{modspec}  for  objects with  different
spectral type  parameters.  The steps we  take are small because  the role of
these  tiny  deviations from  the  template based  spectra  is  to point  the
eigenspectra into the right direction in a statistical sense.

Once  the spectrum  repairing  procedure is  applied  to each  galaxy in  the
training  catalog,  the Karhunen-Lo\`eve  transformation  can  be applied  to
obtain  a  new  set of  eigenspectra.  Repeating  the  above steps  yields  a
statistically  robust training  algorithm to  develop  empirical eigenspectra
from photometry.  If  there were no modifications to  the linear combinations
of the eigenspectra  then we would get exactly the  same spectrum subspace at
the end.  Simply  adding some uncorrelated noise to  the spectra would retain
the subspace as well.  Corrections in  the eigenspectra can only occur due to
changes correlated with galaxy type.

\section{Application to the HDF/NICMOS data} \label{app}
 
\subsection{Building the templates} \label{realtemp}

Our  algorithm  is constructed  in  such  a way  that  the  iteration can  be
initialized with any eigentemplates and started from imported spectra. In our
study for deriving the initial set of eigentemplates we used the extended CWW
spectra, which emerged as the most  commonly used set of templates by several
other groups.

The actual  implementation was somewhat  different. Logarithmically resampled
templates  were  used  in  order   to  speed  up  the  computation,  since  a
logarithmically rebinned spectrum can  be redshifted by simply offsetting the
indices of  the vectors.   The result  of this is  that the  integral measure
changes, which  requires some trivial  modifications in the  above equations.
Having more  bins in  the blue than  the red  ranges (up to  25000\AA) allows
higher resolution in the optical bands.

\begin{figure}
\epsscale{0.5}
\plotone{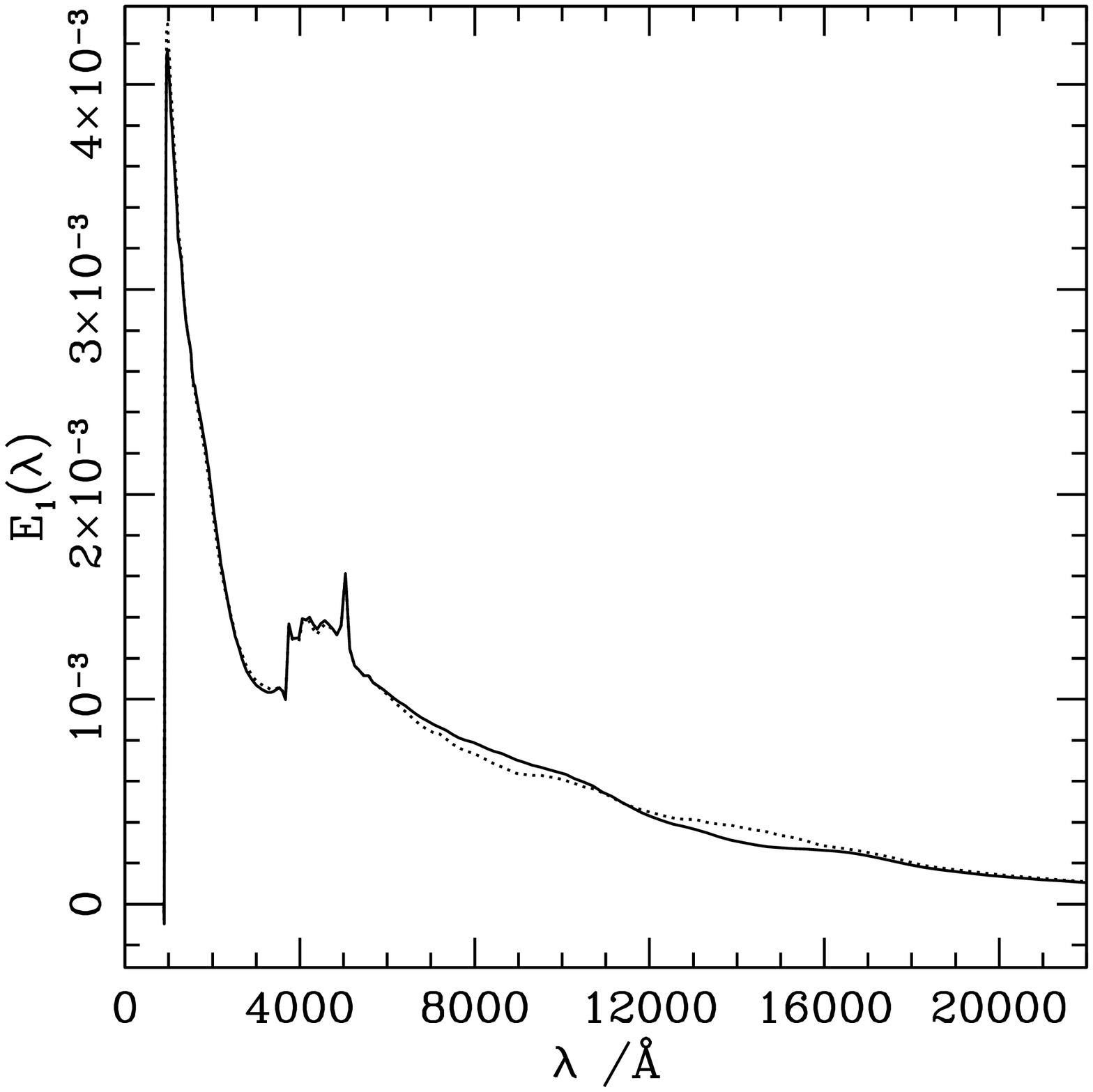}\plotone{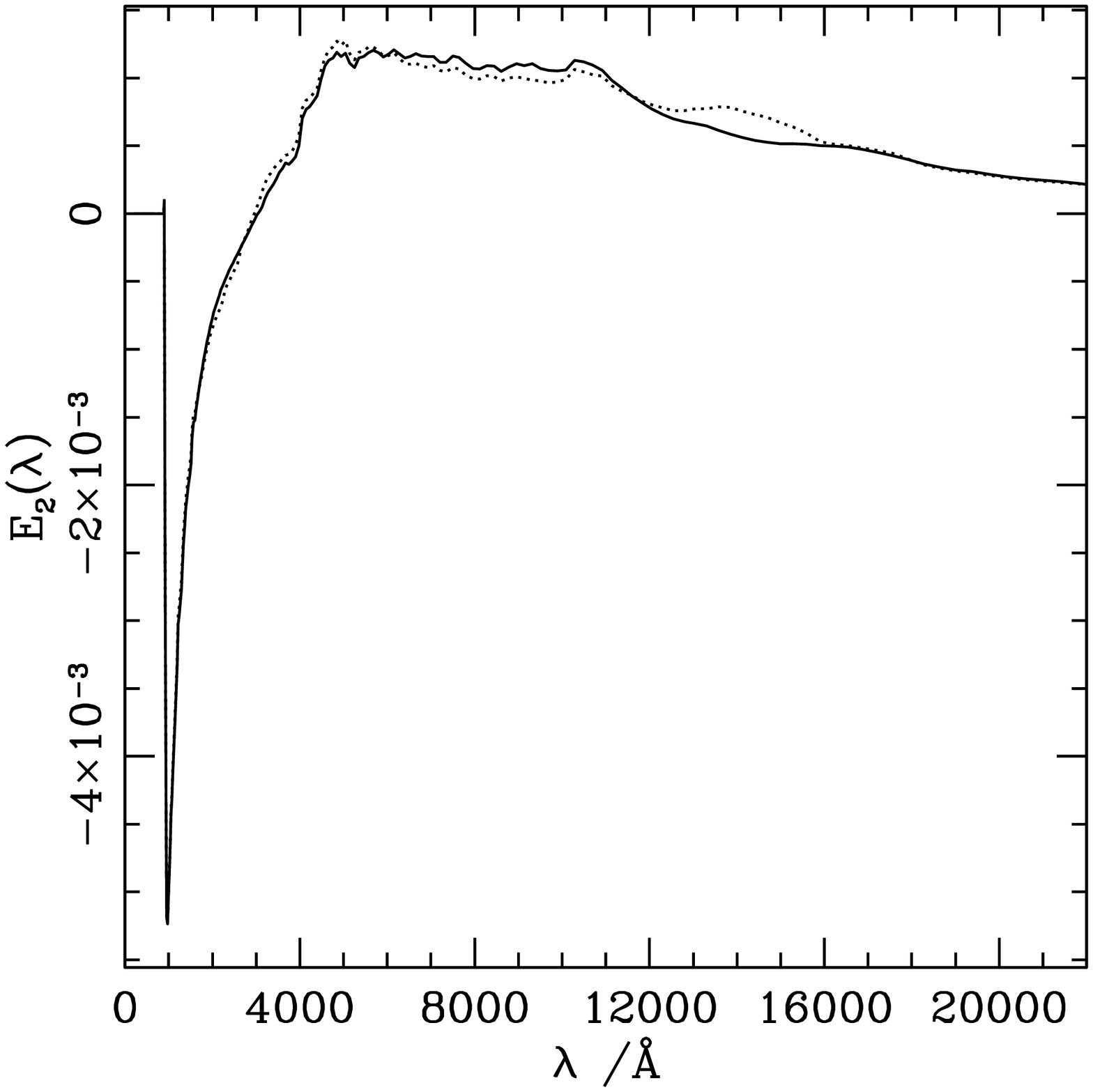}
\plotone{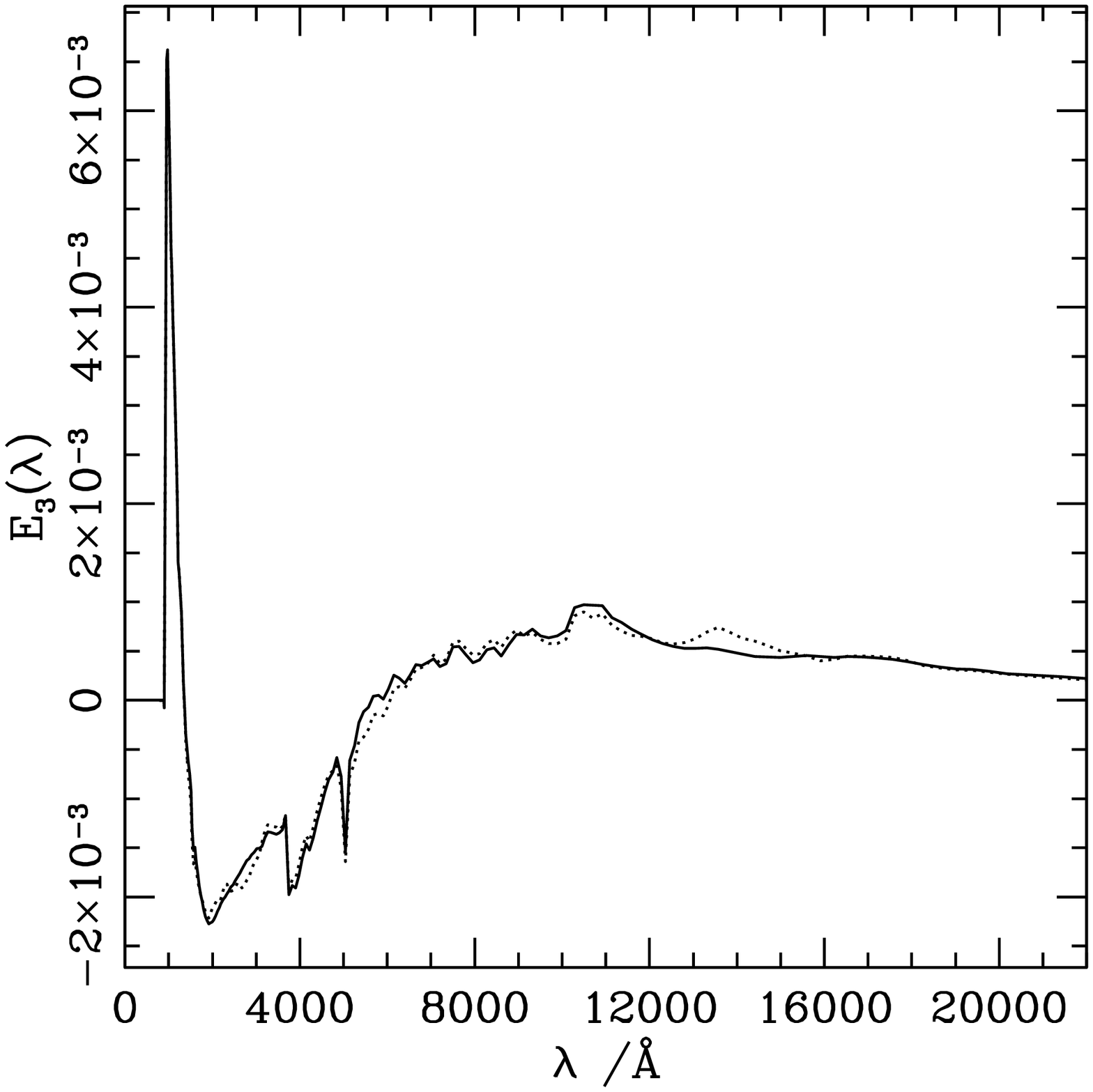}
\caption{The modified eigenspectra (dotted lines)
differ from the original ones (solid lines) derived from CWW spectra. The
figure shows how the overall shapes are adjusted after two iterations. The
eigentemplates are normalized, their dot products give the unit matrix,
$\langle E_i, E_j \rangle = \delta_{ij}$. \label{eifast}}
\end{figure}

\begin{figure}
\epsscale{0.5}
\plotone{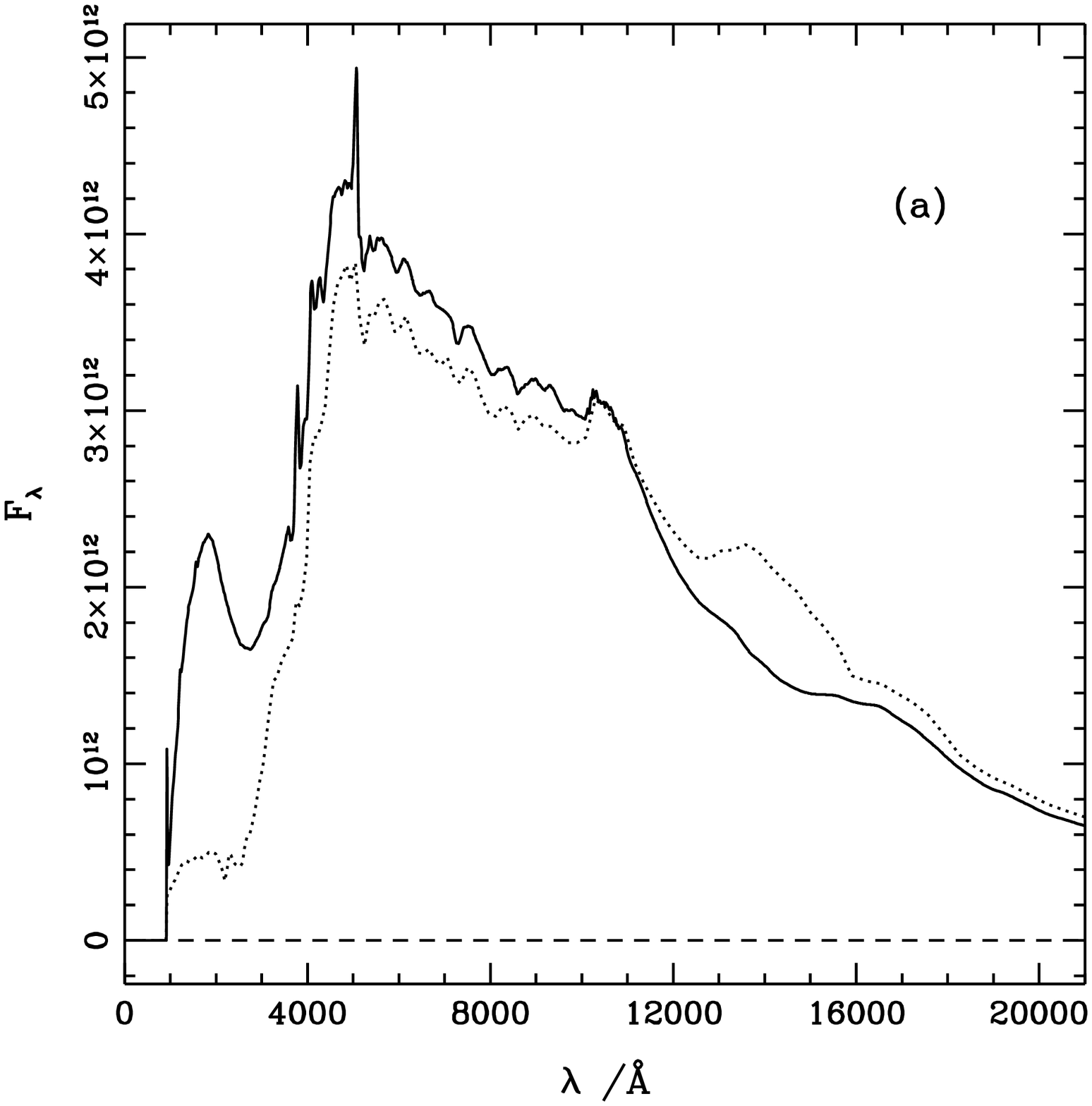}\plotone{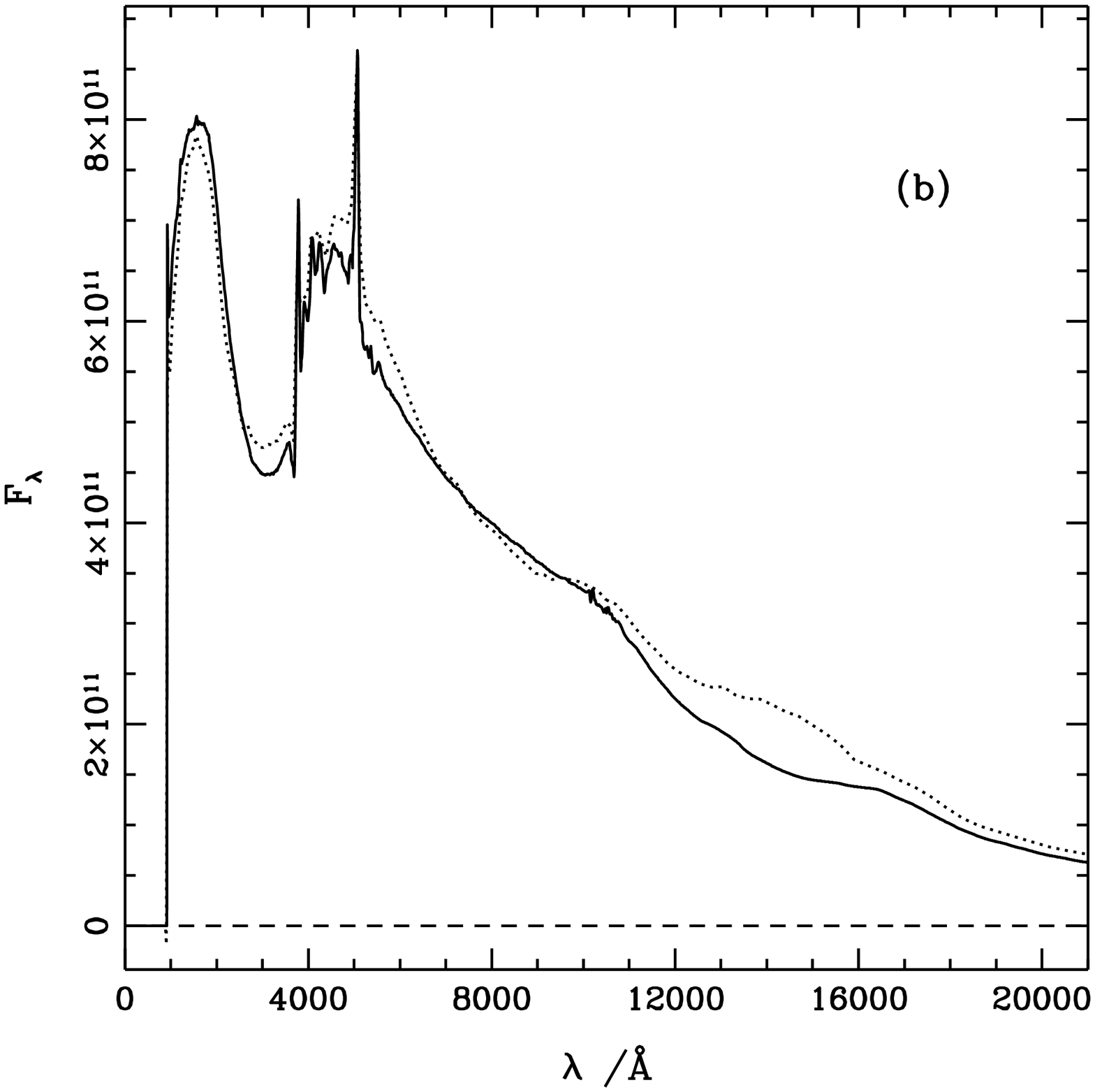}
\plotone{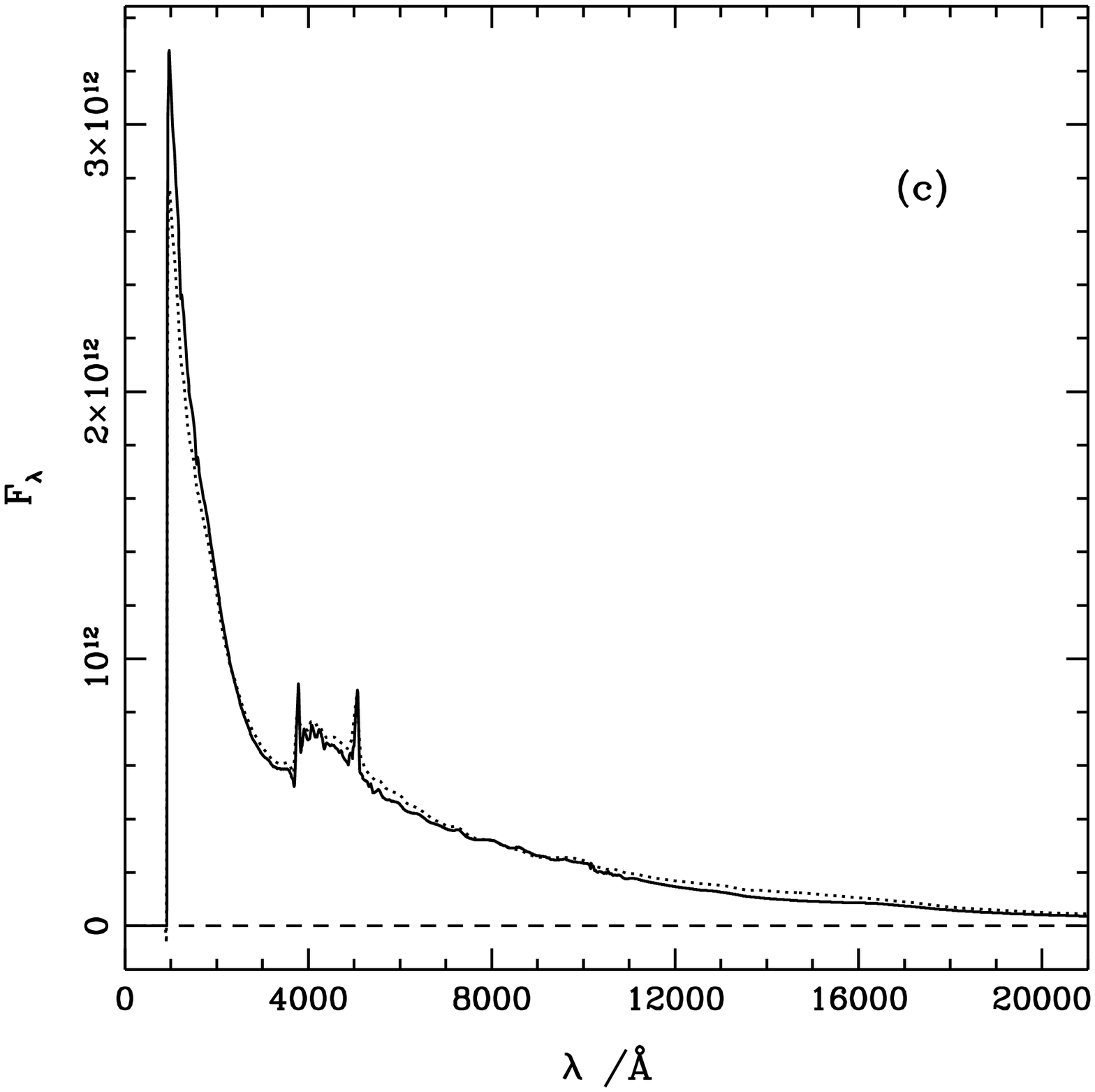}\plotone{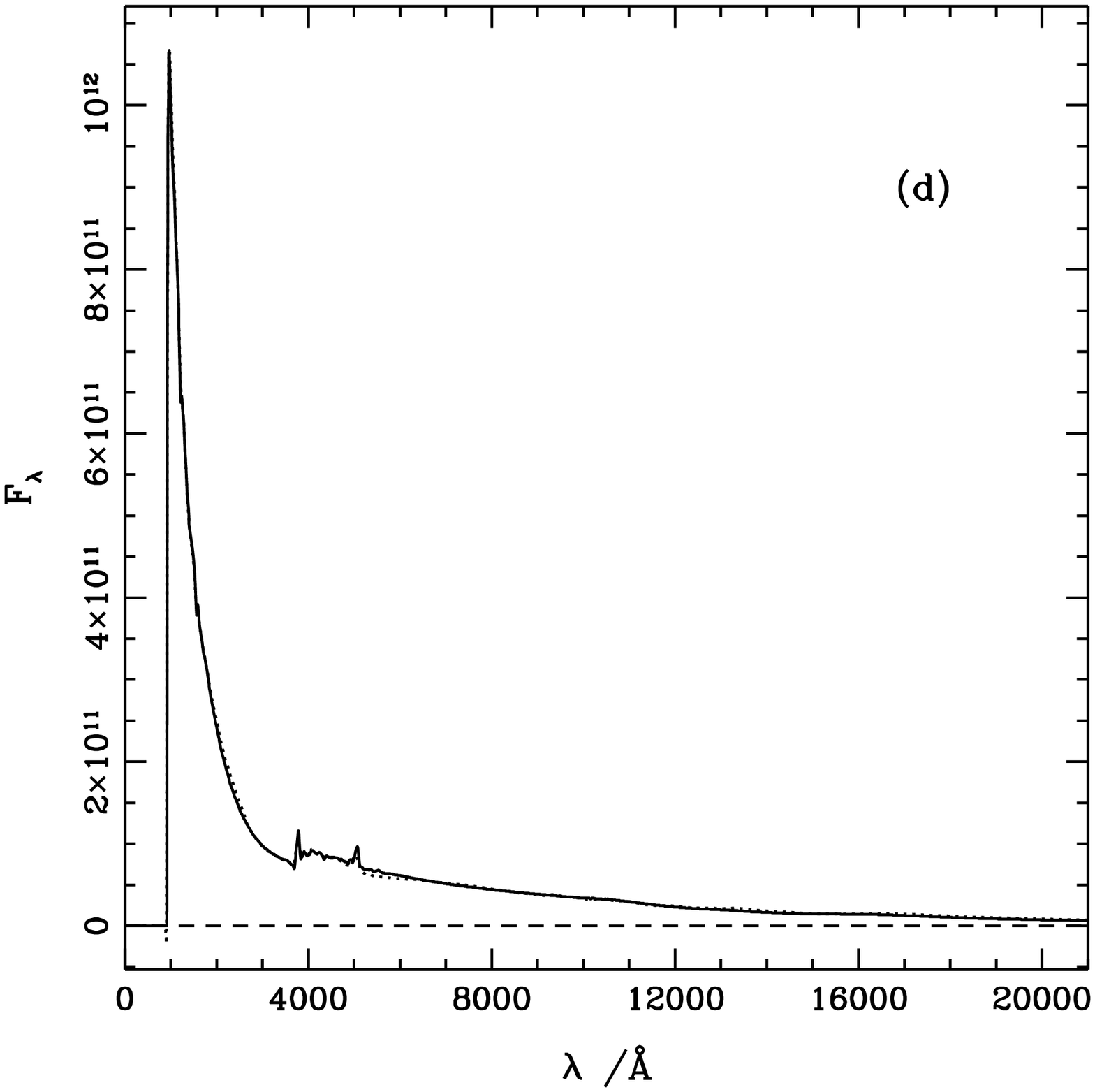}
\caption{Typical estimated spectra differ
significantly especially for early type templates. Both the UV and IR ranges
are modified and templates tend to become redder with the iteration. Sample
objects (NICMOS Id (a) 605, (b) 1023, (c) 1076 \& (d) 561) illustrates the
changes as a function of wavelength (solid line - CWW, dotted - KL[2]). The
estimated redshifts and types for each objects are given in table
\ref{tblcmp}. \label{cmpspec}}
\end{figure}

Repairing the  eigentemplates improved the rest-frame  colors rapidly. Global
features changed significantly over the  initial iterations. After just a few
steps  changes  to  the  continuum  of  the  eigenspectra  converge.   Figure
\ref{eifast} illustrates  the continuum changes  of the basis after  just two
iterations.   Subsequent iterations  change the  higher  frequency components
within  the  eigenspectra.    Figure  \ref{cmpspec}  and  table  \ref{tblcmp}
containing the actual  numeric values of the estimates  help us to understand
the effects  of these  iterations on the  estimated spectra.  Changes  in the
eigenbasis result  in a new fit,  the expansion coefficients  differ and thus
the actual estimated SEDs can differ substantially more than one might expect
from  figure  \ref{eifast}.   Consistent  with  our  expectations  about  the
uncertainties present within the  empirical SEDs, most modifications occur in
the ultraviolet  and the  infra red, where  the initial templates  are poorly
constrained. The bluest starburst SEDs tend to be changed least.

Each iteration requires approximately 2 minutes on a workstation with the CPU
time scaling  linearly with the  number of objects  (as most of the  time was
spent in the  spectrum correction phase solving the  linear equations for the
modified spectrum).  The KL expansion took an insignificant amount of time on
such a small data set.  However, even the principal component analysis can be
done iteratively,  which enables  us to apply  this technique to  much bigger
surveys, such  as the  Sloan Digital Sky  Survey.  Our training  algorithm is
applicable  to  data  sets  composed  of catalogs  in  different  photometric
systems.  Theoretically there  is nothing  to prevent  us from  including all
observed galaxies which have photometry and redshifts available.

\begin{figure}
\epsscale{0.85}
\plotone{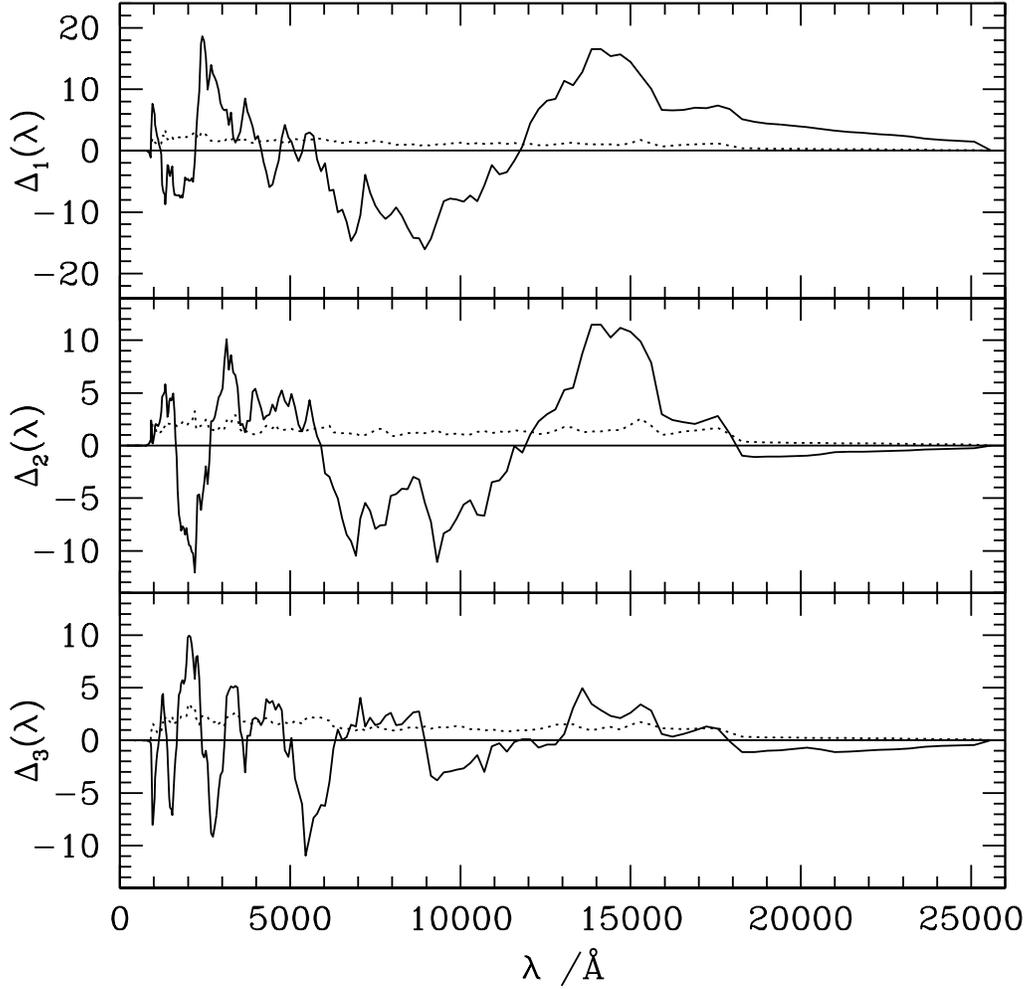}
\caption{Using random subsamples as training sets gives
us the possibility to test the stability of the algorithm.  Having {\it
repaired} the eigentemplates (using KL[10]) on 30 training sets, one can
compute their mean in order to compare them to the original CWW eigenspectra.
Solid lines show the deviation ($\Delta_i$) of these modified templates from
the original basis and the scatter (rms) around the mean spectra plotted with
dotted lines.  The curves are normalized so that the mean scatter (computed
over the entire wavelength range) is unity in each panel. The figure shows
that our corrections in the templates are significant. Panels correspond to
the first, second and third eigencomponents respectively from the top to the
bottom. \label{shuffle}}
\end{figure}

The stability of  our technique had been tested  with randomly chosen subsets
of  the entire training  set. The  iterated eigenbases  were compared  to the
original and the  difference of their projections onto  the same subspace has
been  analyzed. Calculating  the mean  and its  standard deviation  of thirty
resulting templates  in all the wavelength  bins shows us  that these changes
are  significant.   Figure \ref{shuffle}  show  the  mean  difference of  the
resulting and CWW eigentemplates with solid lines and the scatter with dotted
lines. The first eigencomponent changes less but it is the most significant.

\subsection{Applying to photometric redshifts} \label{applphotz}

In our analysis the spectroscopic redshifts  have only been used to shift the
templates to a common rest-frame for the training set. The template repairing
algorithm does not yield photometric  redshifts nor is it directly related to
the estimation at  all.  It reconstructs the (continuum)  rest-frame shape of
galaxy SEDs.  These  eigenspectra can be utilized in  many different ways; to
use  them  as  templates  in  photometric redshift  estimation  is  just  one
possibility.

\begin{figure}
\epsscale{0.37}
\plotone{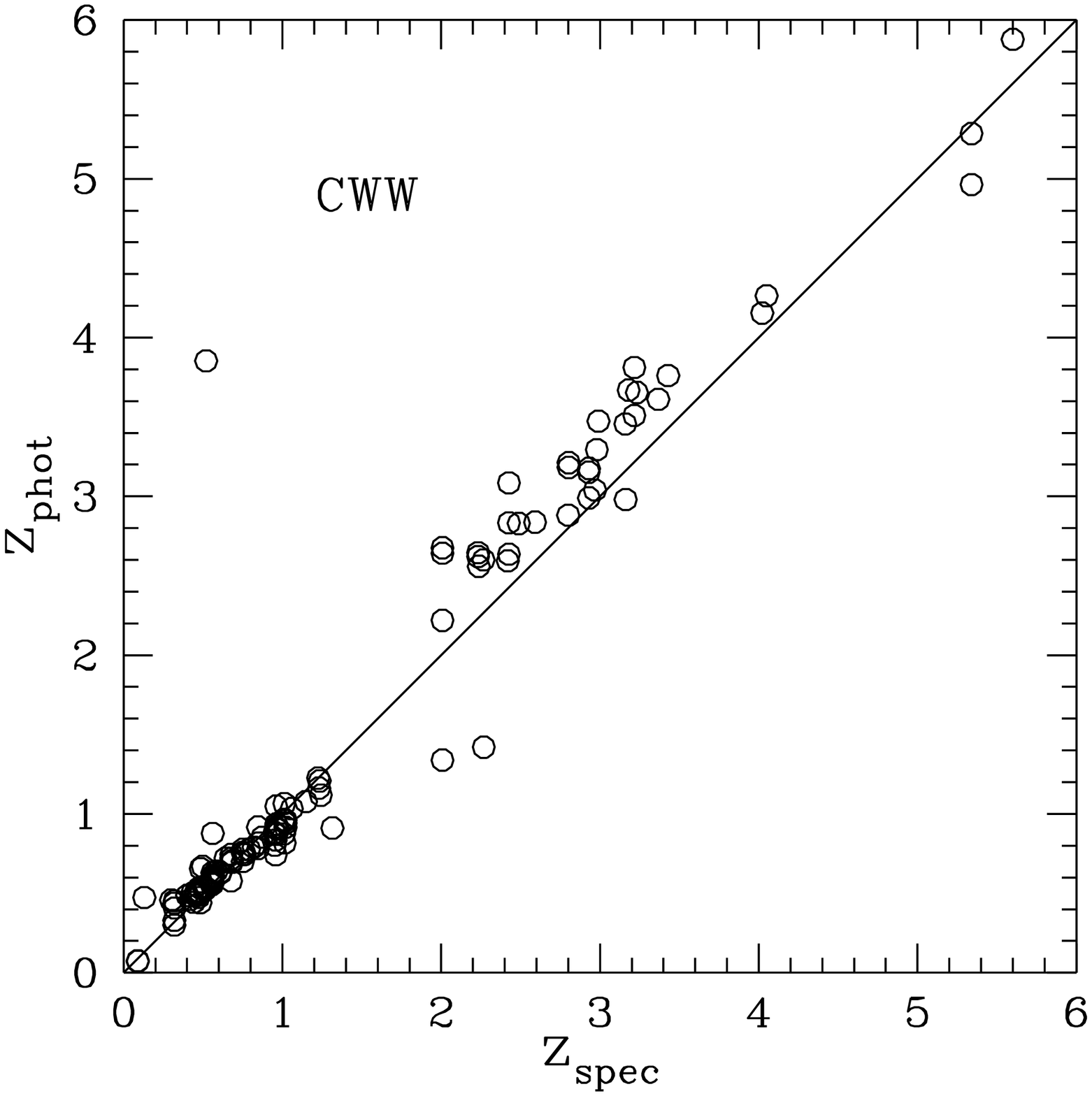}\plotone{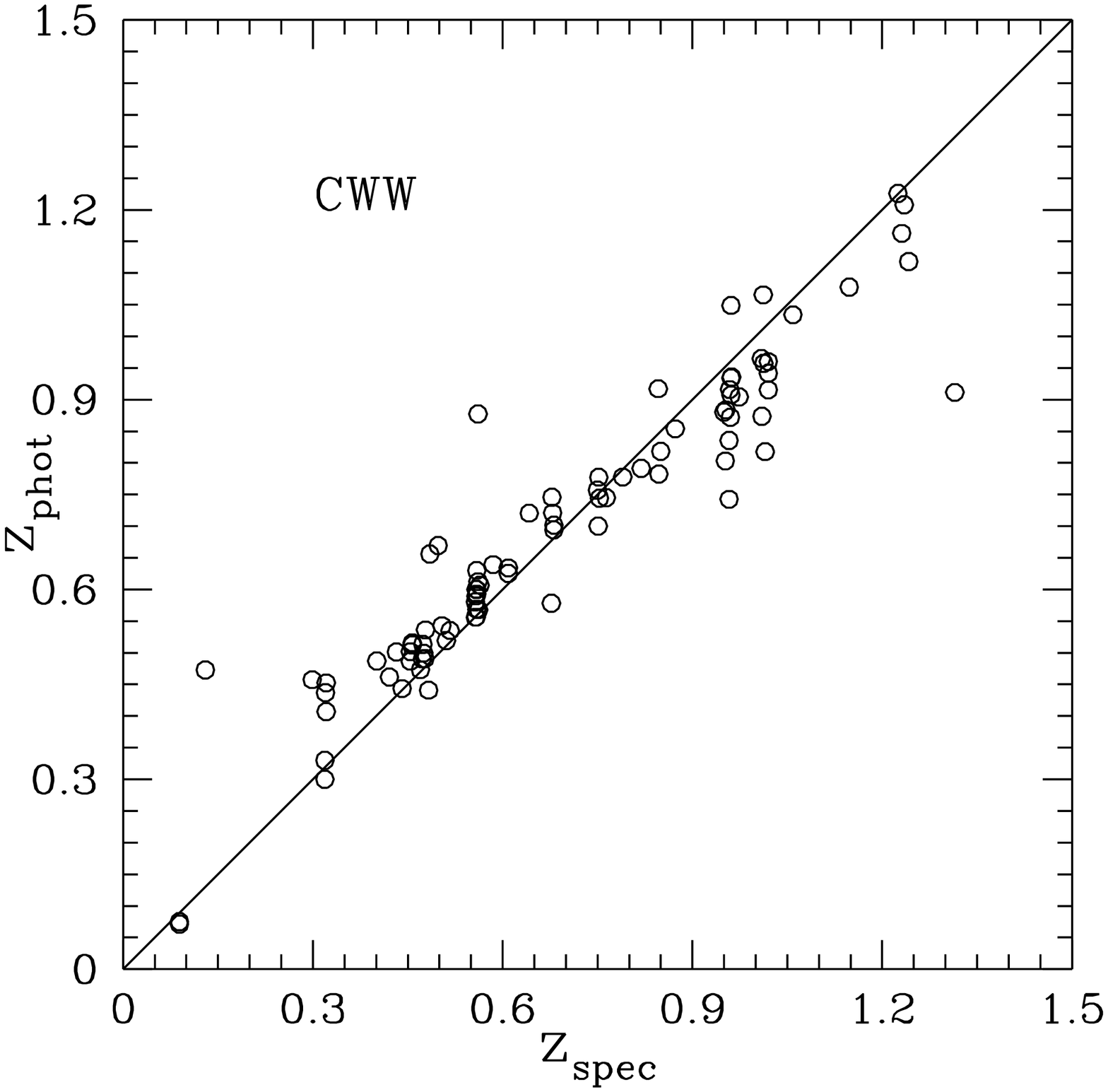}
\plotone{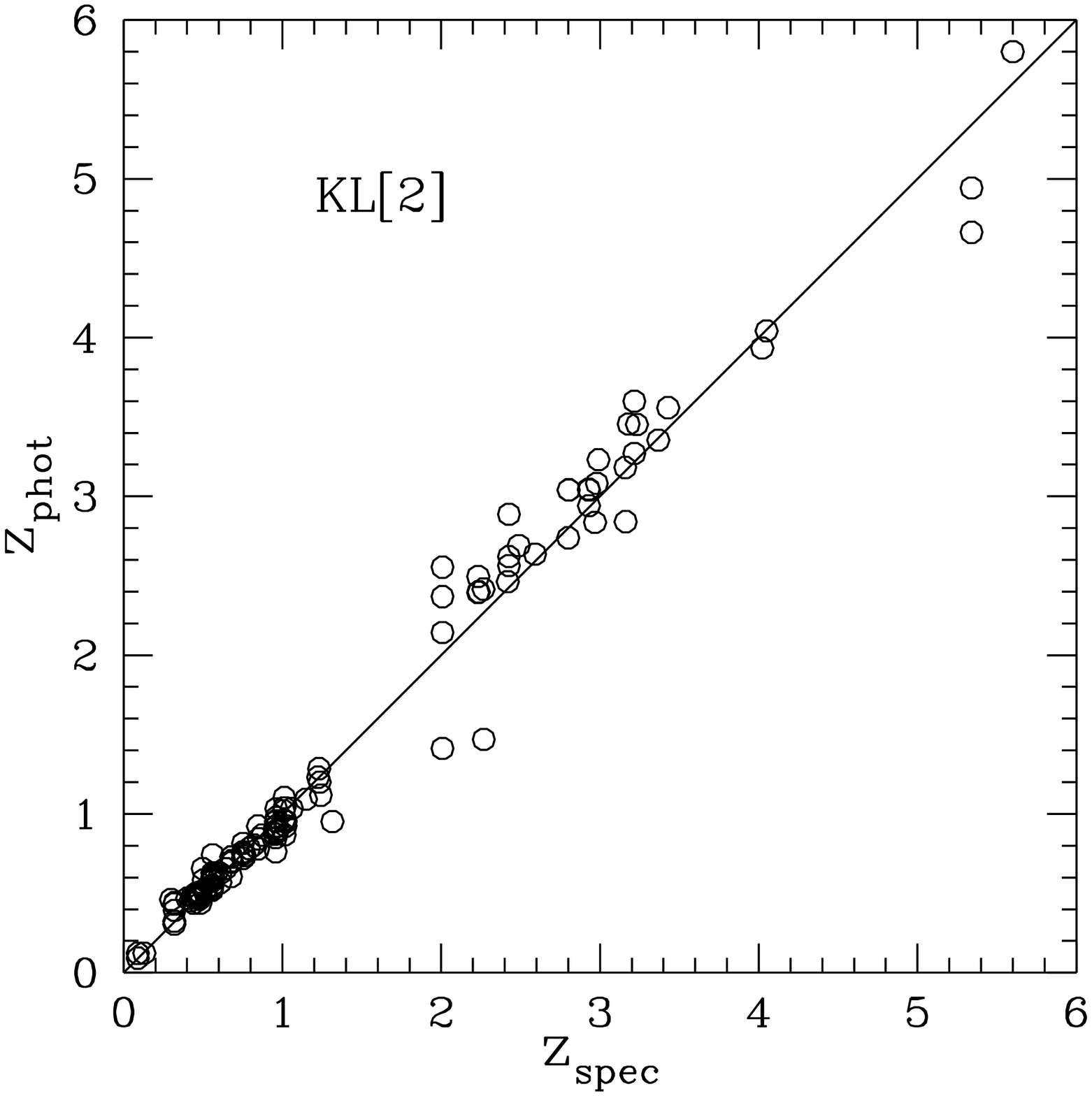}\plotone{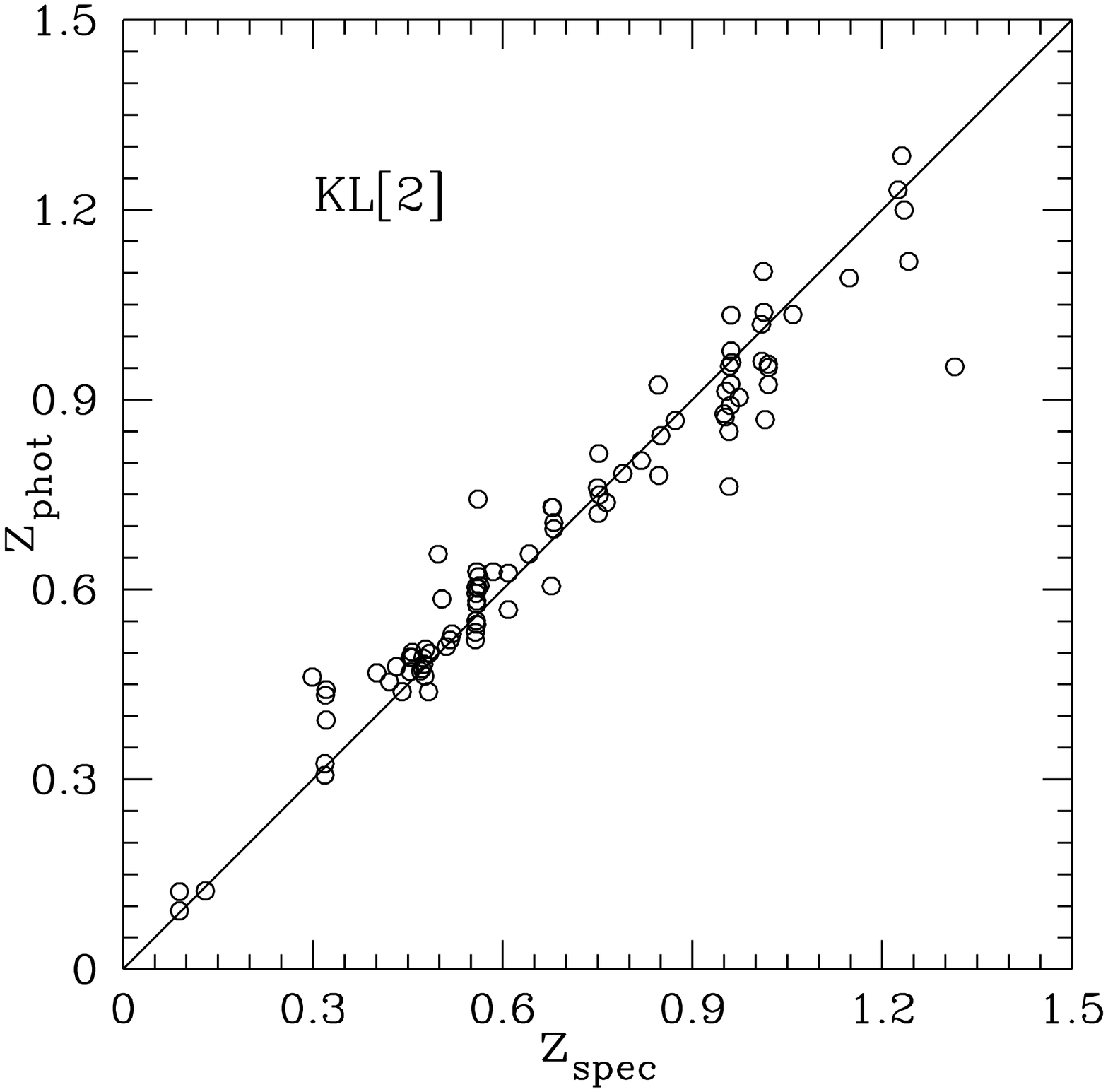}
\plotone{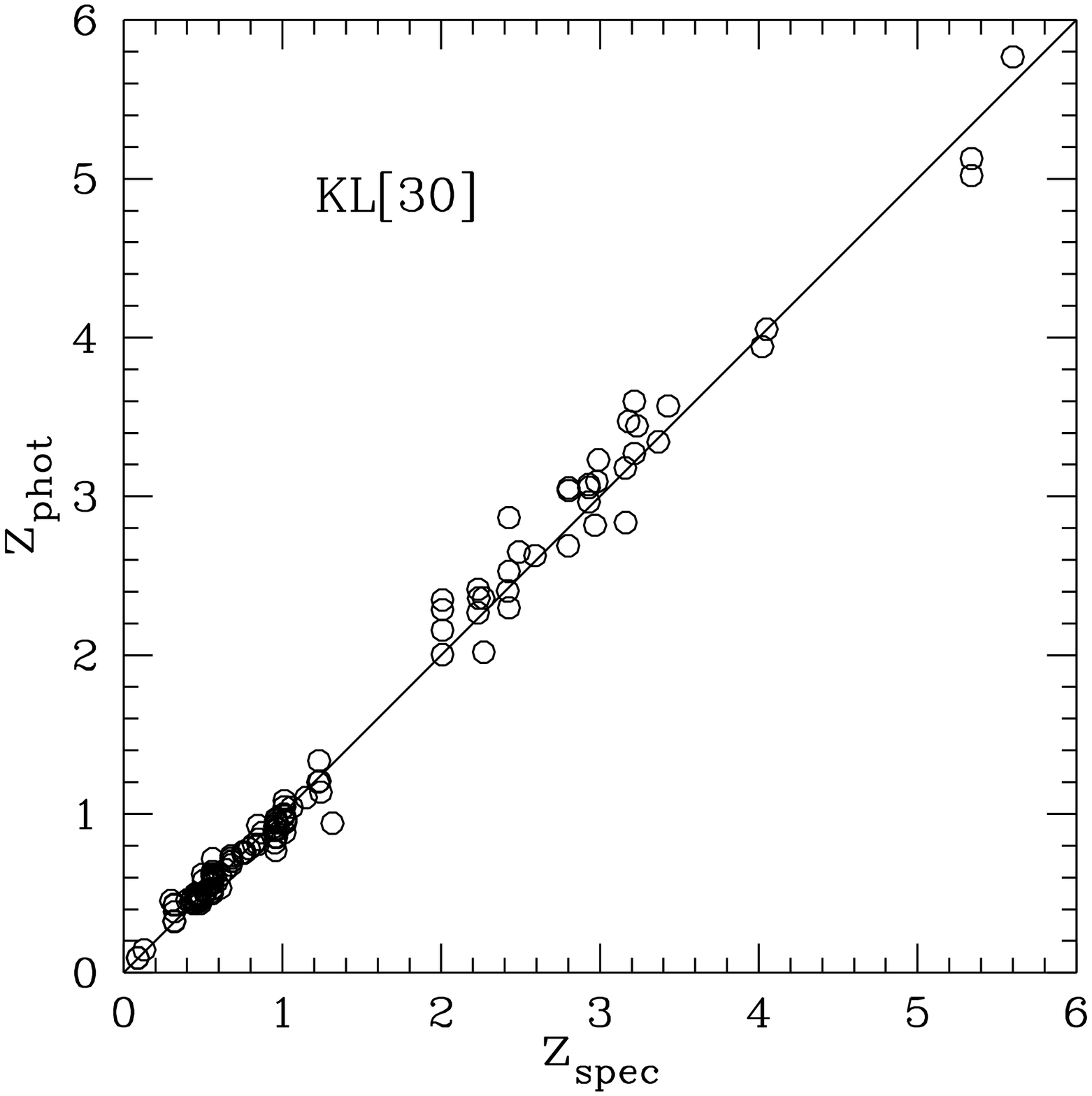}\plotone{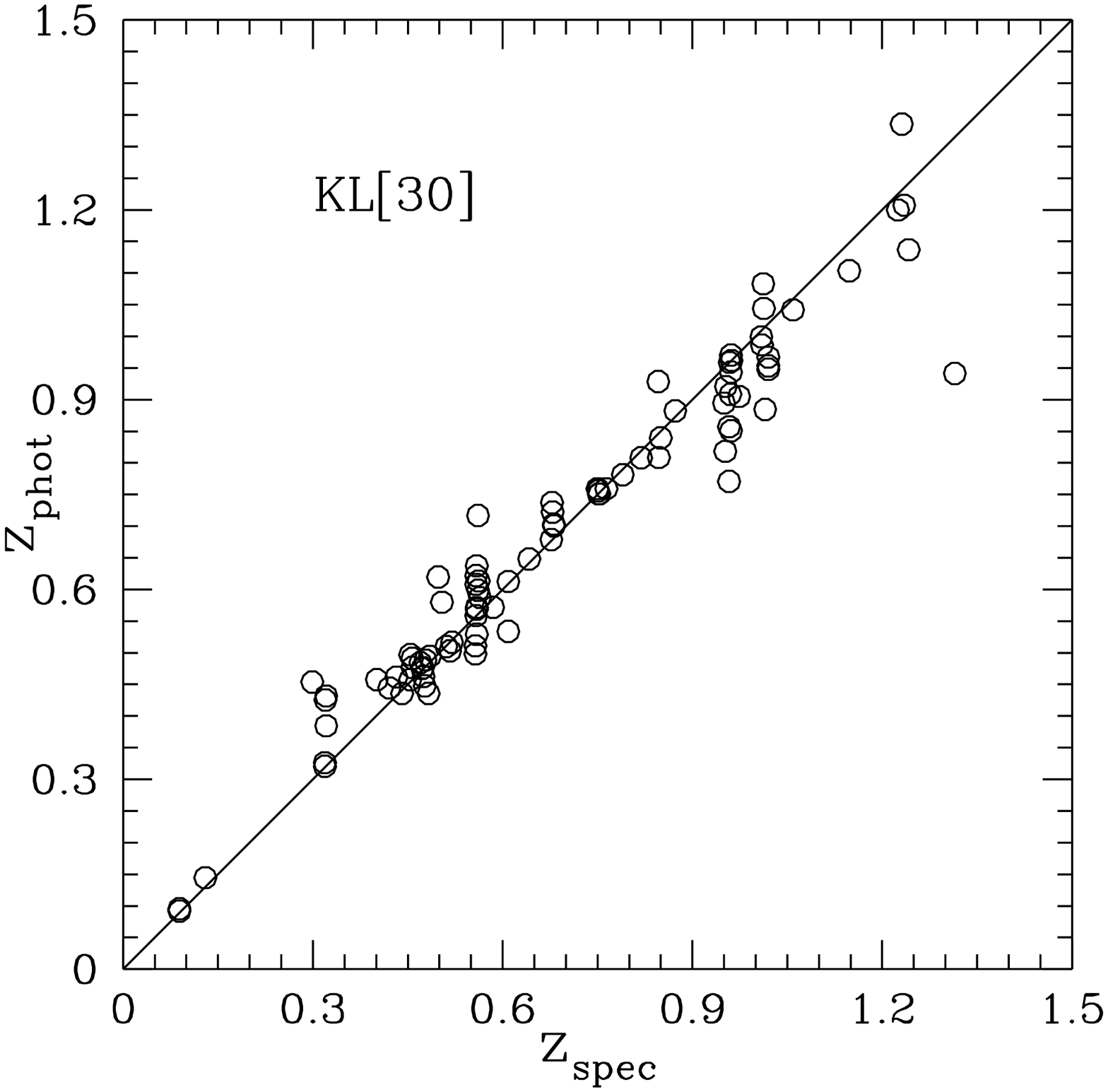}
\caption{The redshift predictions change as the eigenspectra evolve with
successive iterations of the template repair algorithm. From top to bottom,
the plots compare spectroscopic and photometric redshifts based on the CWW,
KL[2], and KL[30] templates. The panels at left show the whole redshift range
available in the HDF, while those at right expand the range $0 < z < 1.5$ for
clarity.
\label{zzKL}}
\end{figure}

Figure  \ref{zzKL}  shows  how   the  redshift  prediction  improves  as  the
eigenspectra  evolve.   The  estimates  are  approaching  the  ideal  $z_{\rm
phot}=z_{\rm  spec}$  line  as  the   SEDs  converge  to  the  best  possible
representation  of  reality.   Figure  \ref{zzKL}  is  a  comparison  of  the
photometric redshift  prediction based on  the resulting KL (bottom)  and CWW
(top)  eigentemplates.   The  systematic  errors disappear  and  the  scatter
reduces.  Beyond the standard root mean square error ($\Delta_{\rm rms}$) the
relative  deviation of  $(1+z)$ was  also computed  ($\Delta_{\rm  rel}$) for
comparison with estimates found by other authors.
\begin{equation}
\Delta_{\rm rms} =
	\left\langle \Big( z_{\rm spec}-z_{\rm phot} \Big)^2 \right\rangle 
\end{equation}
\begin{equation}
\Delta_{\rm rel} =
	\left\langle \Big( {z_{\rm spec}-z_{\rm phot} 
		\over 1+z_{\rm spec}} \Big)^2 \right\rangle
\end{equation}
Table \ref{tblscatter} compares the overall redshift errors computed for both
the CWW and the trained  (KL) eigenspectra. It shows significant improvements
on any redshift range. The final KL eigenbasis gains about a factor of two in
the  average regardless  of which  statistics are  preferred.  The  rms error
changes from 0.23 to 0.12 for estimates up to redshift of 6.

\begin{figure}
\epsscale{0.85}
\plotone{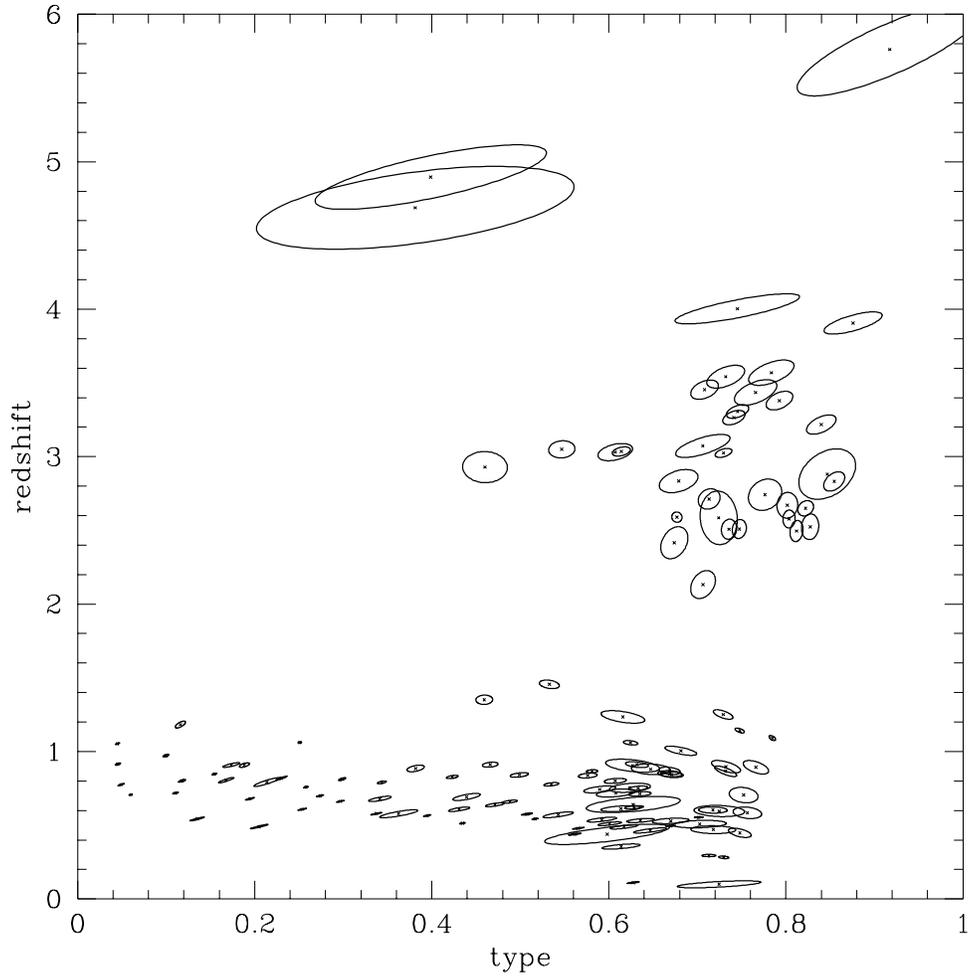}
\caption{Using one-dimensional constraint type fit
allows us to work out the formal covariance matrix of the type and
redshift. The error ellipses show the correlated errors for objects in the
training set. \label{covar}}
\end{figure}

An  important feature of  this approach  is the  rather rigorous  handling of
error  propagation.  The  available photometric  errors were  incorporated in
both the  template repair and photometric redshift  estimation routines.  The
photometric  inversion should predict  not only  the physical  parameters but
also their errors (that are usually correlated). Since it is hard to find the
global  minimum  of a  function  in more  dimensions,  we  evaluate the  cost
function on a grid to speed up the processing and to ensure the optimization
does not end up in a  local minimum.  Searching the grid on the type-redshift
plane provides a good estimate of the global minimum, however the results can
be revised using the continuous  cost function again.  Fitting a second order
surface around the minimum allows us to refine the estimation by choosing the
fitted  minimum  and  to  compute  the  formal  covariance  matrix  from  the
coefficients of the fitting formula.  Figure \ref{covar} shows the correlated
type and redshift errors for the training set.  Consistent with expectations,
most  of the ellipses  show that  higher redshifts  than those  estimated are
possible if  we choose  an incorrect (bluer)  template. This  illustrates the
need to  fully and  finely sample the  distribution of template  spectra. The
bluer SEDs also tend to have larger  errors on the type and redshift than the
ellipticals due to the less significant continuum features in their spectra.

\section{Discussion} \label{discussion}

Our technique has  been shown to be able to  derive rest-frame templates from
photometry in an iterative way, improving step-by-step the correlation of the
template based  colors to the actual measurements.   The statistical approach
of the KL expansion provides a robust determination of the systematic changes
in the  SEDs. After  just the initial  2-3 iterations, the  overall continuum
shapes  of  the  SEDs  are  adjusted to  reproduce  the  photometry.   Higher
resolution detailed features emerge in later iterations.  Consistent with our
expectations, significant continuum changes occur both in the ultraviolet and
infrared, where the original  templates are poorly constrained. The estimated
SEDs based  on the trained eigenbases tend  to be redder in  general.  The UV
part of  the spectra get  fainter and  the flux increases  in the IR  for the
early type  galaxies (which  change the most).   The blue galaxies  have gone
through less dramatic  modulation.  Clearly, the color spread  of galaxies is
larger than one might anticipate from the CWW spectra.

The limitation on  our current application is the  relatively small number of
objects with  accurate multicolor  photometry and spectroscopic  redshifts in
the Hubble Deep  Field. Combining several observations into  one training set
is  one way  to improve  the  robustness of  our results.   Our technique  is
constructed to be able to deal  with many multicolor observations at the same
time. Data of  various surveys can be incorporated  regardless of photometric
systems they use, meaning not just broad but also narrow or intermediate band
imaging.  This  flexibility of the algorithm  allows us to  simply merge data
sets coming  from several instruments, where  the data set  up constraints on
the SEDs  at different  wavelengths.  With  time we will  be able  to improve
other  parts  of  the  SEDs.   New upcoming  instruments  will  provide  high
signal-to-noise  ratio data in  different spectral  ranges, e.g.   the Galaxy
Evolution   Explorer   (GALEX)  probing   the   far   and  near   ultraviolet
(1300-3000\AA) or the Space Infrared  Telescope Facility (SIRTF) and the Next
Generation Space Telescope  (NGST) looking at the infrared  ranges.  In fact,
the   algorithm  is  even   more  generic.    Eventually,  spectrophotometric
observations can be naturally incorporated in the training mechanism.

There remains  one class of galaxy  for which this technique  is not optimal;
high  redshift  galaxies for  which  we  cannot  easily obtain  spectroscopic
redshifts can  not be optimized by  our procedure (i.e. we  lack the relevant
training set).  As our technique  can seamlessly incorporate  theoretical and
empirical data (weighting the samples  appropriately) if we have a reasonable
understanding of  the stellar populations  we can account for  missing galaxy
populations. Objects of extremely unusual  colors will not be well matched by
any template  fitting procedure but  the goodness of  fit to these  data will
enable us to identify these anomalous objects.


In this  paper we have presented  a new method to  reconstruct the rest-frame
continuum spectra of galaxies directly from a catalog of broadband multicolor
photometric  observations and  spectroscopic  redshifts.  In  addition to  an
improved photometric  redshift estimation the application  of eigenspectra to
this  problem enables the  classification of  galaxy spectra  to move  from a
series of discrete (and somewhat ad hoc) classification types to a continuous
parameterization.   With   the  help  of  the   optimal  subspace  filtering,
non-physical spectral energy  distributions could be avoided. As  a result of
the repair algorithm, the eigenbasis improved the capability to represent the
observed photometric properties. This  improvement cannot be tested directly,
since we  do not  know the real,  photometrically calibrated spectra  for the
objects in our training catalog, but the photometric redshift estimation test
with  the  repaired  spectra  gave  substantially  better  results  than  the
estimation   using   model/empirical   templates   ($\Delta_{\rm   rms}=0.12$
vs. $\Delta_{\rm rms}=0.23$). New  photometric and spectroscopic surveys will
extend the size  of the training catalog, making  possible to improve further
the spectral templates and enrich them with finer details.

\acknowledgements 

IC and TB acknowledges partial support from the MTA-NSF grant no. 124
and the Hungarian National Scientific Research Foundation (OTKA) grant
no.\ T030836, AJC acknowledges support from an LTSA grant (NAG57934)
and from NASA through grant number GO-07817.06-96A from the Space
Telescope Science Institute.  AS acknowledges support from NSF
(AST9802980) and a NASA LTSA (NAG53503).  MD received support from
NASA through grant number GO-07817.01-96A from the Space Telescope
Science Institute, which is operated by the Association of
Universities for Research in Astronomy, Inc., under NASA contract
NAS5-26555.

\newpage

\begin{deluxetable}{rcrrrrrr}
\tablecolumns{8}
\tablewidth{0pc}  
\tablecaption{Comparison  of  estimated redshifts  and  types  of  sample
objects shown in figure \ref{cmpspec}.}
\tablehead{
\colhead{HDF} & \colhead{NICMOS} & \colhead{} &
\multicolumn{2}{c}{CWW} & \colhead{} &
\multicolumn{2}{c}{KL[2]} \\
\cline{4-5} \cline{7-8}
\colhead{Id} & \colhead{Id} & \colhead{$z_{\rm spec}$} & 
\colhead{$z_{\rm est}$} & \colhead{$t_{\rm est}$} & \colhead{} &
\colhead{$z_{\rm est}$} & \colhead{$t_{\rm est}$}}
\startdata
1-34.1  & 605  & 0.485 & 0.66 & 0.16 & & 0.50 & 0.05 \\ 
4-173.0 & 1023 & 0.959 & 0.91 & 0.38 & & 0.95 & 0.35 \\ 
4-888.0 & 1076 & 1.010 & 0.87 & 0.68 & & 0.96 & 0.64 \\ 
2-824.0 & 561  & 2.419 & 2.54 & 0.84 & & 2.48 & 0.84 \\
\enddata
\label{tblcmp}
\end{deluxetable}

\begin{deluxetable}{crrrrrrr}
\tablecolumns{8}
\tablewidth{0pc}  
\tablecaption{The accuracy of the redshift prediction is compared for the
CWW, KL[2] and KL[30] eigenbases. Both the rms and relative errors improve
significantly with the training.}
\tablehead{
\colhead{Redshift} & 
\multicolumn{3}{c}{$\Delta_{\rm rms}$} & \colhead{} &
\multicolumn{3}{c}{$\Delta_{\rm rel}$} \\
\cline{2-4} \cline{6-8}
\colhead{range} & 
\colhead{CWW} & \colhead{KL[2]} & \colhead{KL[30]} & \colhead{} &
\colhead{CWW} & \colhead{KL[2]} & \colhead{KL[30]}}
\startdata
$z<1.5$   & 0.096 & 0.073 & 0.070 & & 0.060 & 0.041 & 0.038 \\
$1.5<z<6$ & 0.391 & 0.293 & 0.197 & & 0.111 & 0.081 & 0.052 \\
$z<6$     & 0.227 & 0.169 & 0.121 & & 0.079 & 0.056 & 0.043 \\
\enddata
\label{tblscatter}
\end{deluxetable}


\end{document}